\documentclass[journal,twoside]{IEEEtran}

\IEEEoverridecommandlockouts

\usepackage{amsmath,amssymb,amsfonts,amsthm}
\usepackage{textcomp}
\usepackage{xcolor}
\usepackage{graphicx}
\usepackage{cite}
\usepackage{color}
\usepackage{url}
\usepackage{balance}
\usepackage{soul}
\usepackage{enumitem}
\definecolor{darkB}{RGB}{0,51,153}
\setstcolor{darkB}

\usepackage{siunitx}
\newtheorem{assumption}{Assumption}

\newtheorem{Prop5}{Theorem}

\newtheorem{Remark}{Remark}

\usepackage{multicol}
\usepackage{multirow}
\usepackage{caption}
\usepackage{makecell}
\usepackage[subtle]{savetrees}
\usepackage{float}
\begin{document}


\title{\vspace{-0.07in}
 Large-Signal Stability Guarantees for a DC Microgrid \\ with Nested Nonlinear Distributed Control:\\ The Slow Communication Scenario}
\author{Cornelia Skaga,~\IEEEmembership{Graduate Student Member,~IEEE},\\ Mahdieh S. Sadabadi ,~\IEEEmembership{Senior Member,~IEEE}, and Gilbert Bergna-Diaz,~\IEEEmembership{Member,~IEEE}
\vspace{-0.15in}

\thanks{The preliminary results of this paper are presented in \cite{Poppi_SEST}. 

Cornelia Skaga with the Department of Electric Energy, Norwegian University of Science and Technology (NTNU), 7491 Trondheim, Norway (e-mail: cornelia.skaga@ntnu.no). 

Mahdieh S. Sadabadi is with the Department of Electrical and Electronic
Engineering, The University of Manchester, M13 9PL, Manchester, UK (e-mail:
mahdieh.sadabadi@manchester.ac.uk).

Gilbert Bergna-Diaz is with the Department of Electric Energy, Norwegian University of Science and Technology (NTNU), 7491 Trondheim, Norway (e-mail: gilbert.bergna@ntnu.no).}
}
\markboth{Submitted for Publication}{Submitted for Publication}

\maketitle

\begin{abstract}
The increasing integration of renewable energy sources into electrical grids necessitates a paradigm shift toward advanced control schemes that guarantee safe and stable operations with scalable properties. Accordingly, this paper investigates large-signal stability guarantees for cyber-physical DC microgrids employing a \emph{nonlinear} distributed consensus-based control scheme to enable coordinated integration and management of distributed generation units within an expandable framework. The proposed control framework adopts nested control loops--inner/decentralized and outer/distributed--specifically designed to simultaneously achieve uniform voltage containment within pre-specified limits, and proportional current sharing in steady state. Our \emph{scalable} stability result relies on singular perturbation theory and Lyapunov arguments to prove global exponential stability when imposing a sufficient time-scale separation at the border between the nested control loops, while relying on some \emph{practical} parameter-setting schemes. The effectiveness and versatility of the proposed control strategy are then validated through time-domain simulations performed on a case-specific low-voltage DC microgrid and the modified IEEE 33-bus radial distribution system. Moreover, a small-signal stability analysis is conducted to derive \emph{practical guidelines} that enhance the applicability of the method.


\end{abstract}


\section{Introduction}
Modern power systems are strategically evolving into converter-dominated, multi-agent grids to enable profitable and efficient integration of distributed (renewable) energy resources (DER) such as photovoltaics, wind turbines, battery energy storage systems, and electric vehicles \cite{Babak_AC, Poppi_SEST}. Furthermore,  the inherent electrical nature of most renewable energy sources (RES), along with modern generation and load technologies, positions multi-agent DC microgrids (MGs) as a promising solution for optimizing local power utilization of these DER, ascribed to their high flexibility and operational advantages to operate in both \emph{grid-connected} and autonomous modes of operations (\emph{islanded}) \cite{Detection_2}, as well as their simplified dynamics, control, and management \cite{Babak_Set_Point}. 
Traditionally, power grids are controlled in a centralized manner, with a single control unit gathering data from the wide-area power grid network. Centralized frameworks often require a sparse communication network, are easy to implement, and are often able to obtain globally optimized solutions \cite{Scalable_Review}. However, as the centralized controller is highly dependent on global system information, the framework is highly susceptible to single-point-of-failure (SPoF) and communication-related disturbances such as communication delays, data-link failures, and data packet losses, while scalability also remains a challenge \cite{Scalable_Review, Event_2, Detection_5}. Hence, the control configurations of modern multi-agent grids -- interconnecting an expanding number of DER -- are undergoing a paradigm shift toward communication-reliant distributed control schemes to enhance flexibility, scalability, and reliability \cite{Babak_Base_Article, 21, Distributed_Overview}. Distributed control strategies often include primary local control of the power electronic converters for local stabilization of the distributed generation units (DGs), and a secondary distributed controller-- leveraging high bandwidth communication between neighboring agents to cooperatively restore initial conditions \cite{Review_Cite_MS_1}. 

However, these converter-dominated DC grids exhibit low electrical inertia and weak damping, making them vulnerable to transient disturbances and less robust against random perturbations, underscoring the need for scalable control strategies that ensure sufficient stability guarantees \cite{new_LR_lyap3, modeling+stability_rev}.

\subsection{Research Gap}
From a performance perspective, this growing integration of DER introduces challenges related to precise voltage stability and power flow management \cite{SS_CS_Book_ch_6}. As a result, proportional load sharing (current- or power sharing) among DGs, along with stable and precise voltage regulation, are typically considered as the primary control objectives in DC microgrids \cite{Distributed_1, Detection_5}. Voltage regulation is essential for ensuring the proper operation of interconnected loads, commonly achieved by maintaining the average voltage across the MG at a global set-point value \cite{Distributed_1}, and load sharing is gaining interest as it enhances scalable capabilities \cite{SP_6}. Conventionally, adapting hierarchical control schemes (such as secondary distributed control frameworks) facilitates the achievement of both objectives simultaneously \cite{Distributed_1}. Several distributed communication-based (cooperative) control configurations for DC MGs have been proposed in the literature \cite{cons_DC_1, cons_DC_2, cons_DC_3, cons_DC_4, SP_5, SP_6, SP_7, Distributed_1}, demonstrating the feasibility of achieving simultaneous current sharing and average voltage regulation. However, average voltage regulation may lead to significant deviations in individual generator voltages, potentially exceeding operational limits \cite{Babak_Set_Point}. To overcome these limitations, \cite{Babak_Set_Point} introduces a novel nonlinear distributed control framework, relying on nested control loops-- inner/decentralized/primary and outer/distributed/secondary-- to ensure that all energized units contribute power proportionally to their rated capacities while dynamically operating within predefined voltage limits. However, the nonlinear control framework was proposed without any guarantee of stability. Lyapunov’s direct method -- commonly used in large-signal stability analysis of nonlinear systems-- can indeed be used to assess the stability of the case-specific system with a defined (and preferably low) number of generators admitting the structure given in \cite{Babak_Set_Point}. That being said, as the proposed control framework employs nested control loops, obtaining a scalable Lyapunov candidate of the general n-generator case becomes nontrivial, thereby limiting its scalability in practice.
Moreover, to address the challenges posed by nonlinear nested control loops, \cite{Popov} employs the \emph{Popov} multiplier in conjunction with Lyapunov theory to ensure large-signal stability of a stand-alone converter. Preliminarily, the approach relies on reformulating the closed-loop system into a suitable \emph{lure-like} representation isolating the static nonlinearity. This facilitates the derivation of dissipativity-based conditions under which Lyapunov stability can be established. Even though the result in \cite{Popov} is a non-conservative large-signal stability certificate, its current formulation seems to be case-specific. 

\subsection{Contributions}
To derive a large-signal stability certificate that ensures \emph{scalable} Plug-and-Play (PnP)-properties while addressing the nonlinear nested control loops in \cite{Babak_Set_Point}, this paper reformulates the system as a singularly perturbed problem with sufficient time-scale separation, and applies Lyapunov theory to establish Global Exponential Stability (G.E.S.). This approach is motivated by the methodology presented in \cite{Babak_AC}, where an almost identical control strategy was proposed for AC grids to ensure scalable global asymptotic stability conditions. That proof relied on singular perturbation theory (SPT), accommodating scalable stability conditions under reasonable time-scale assumptions for AC grids; i.e., the distributed outer-loop exhibits fast behavior while the electrical system operates at a slower time-scale.  Moreover, our preliminary work in \cite{Poppi_J1} extends the theoretical foundation in \cite{Babak_AC} to the DC grid in \cite{Babak_Set_Point}, guaranteeing scalable G.E.S. under similar time-scale assumptions. However, given the commonly adopted hierarchical control architectures in DC microgrids, it arguably more reasonable to assume that the outer-loop dynamics (distributed communication-based) operates at a slower time-scale than the inner-loop dynamics (decentralized integral controller and electrical dynamics). Accordingly, to facilitate expressing the system as a singularly perturbed problem under this time-scale separation, and subsequently employ Lyapunov arguments to prove G.E.S., we modify the proposed distributed control strategy in \cite{Babak_Set_Point} and advance the result originally introduced in \cite{Poppi_SEST}. First, we update our control dynamics in the following ways; 1) we saturate the influence of the distributed outer-loop controller in the decentralized inner-loop to preserve a more convenient mathematical structure, 2) we include a proportional term to improve performance, leading to a PI-like control structure, 3) we introduce a constant leakage in the primary communication state to facilitate large-signal stability guarantees, 4) we include a constant leakage in the decentralized integral controller for the small-signal stability analysis. Secondly, the stability proof has been substantially updated to accommodate these modifications, followed by some \emph{practical} parameter-setting guidelines to guarantee G.E.S. and proportional current sharing under both saturated and unsaturated operations. Furthermore, the case studies have been extended to empirically validate the theoretical assumptions and parameter settings, as well as to assess the control performance in continuously ensuring voltage containment and convergence to a steady state satisfying proportional current sharing. The control framework has also been tested on a radial distribution benchmark system to verify its general applicability and scalability. Finally, we conduct a small-signal stability analysis to assess the potential for reducing the conservativeness of the large-signal stability result. 

The rest of the paper is structured as follows. Section \ref{Model} presents the model of the electrical system and the proposed control framework. In Section \ref{Stability}, we present the main result of this paper: a comprehensive stability proof for a closed-loop DC microgrid with a nonlinear distributed controller, that under certain conditions can stabilize to a desired (near-optimal) steady-state. Furthermore, in Section \ref{sec:tuning} we propose some \emph{practical} parameter settings to guarantee stability, followed by Section \ref{sec_Case_Studies} in which the effectiveness of our proposed method is tested through time-domain simulations on a 4-terminal DC microgrid and a 33-bus radial distribution system, complemented by a small-signal stability analysis. Finally, Section \ref{conc} concludes this paper.  
\section{System Modeling and Control Framework} \label{Model}
Throughout the paper, we use the following notations: $\mathbb{R}^{n\times m}$ and $\mathbb{R}^n$ denote a set of $n\times m$ real matrices and $n \times 1$ real vectors, respectively. $\mathrm{col}(\cdot\cdot\cdot) \in \mathbb{R}^n$ denotes a column vector and $\mathrm{bcol}\{\cdot\cdot\cdot\}\in \mathbb{R}^n$ denotes a column vector of vectors. $\mathrm{diag}(\cdot\cdot\cdot)\in \mathbb{R}^{n\times n}$ denotes a diagonal matrix with scalar entries, and $\mathrm{bdiag}\{\cdot\cdot\cdot\}\in \mathbb{R}^{n\times n}$ denotes a diagonal matrix of vectors. $0_n\in \mathbb{R}^n$ denotes a null vector, $0_{n\times n} \in \mathbb{R}^{n \times n}$ denotes a null matrix, and $1_{n\times n}\in \mathbb{R}^{n \times n}$ denotes the identity matrix of appropriate dimensions. $A|_\mathrm{sym}$ denotes the symmetrical part of an arbitrary square matrix $A \in \mathbb{R}^{n \times n}$, defined as $A|_\mathrm{sym}=\frac{1}{2}(A+A^\top)$. Furthermore, $x \gg y$ denotes that x is much larger than y, and $\mathbb{R}_{\geq 0}$ defines all non-negative real values. Given a scalar or a vector $x$, the value at the equilibrium point is indicated as $\Bar{x}$, and $\tilde x$ denotes a shifted variable where $\tilde x \triangleq x - \bar x$. Moreover, we use the block vector notation $x^b$, for any subset $b \subseteq \mathcal{G} \cup \mathcal{E} \cup \mathcal{N}$, where $x^b$ denotes the vector collecting the states associated with the indices in $b$. Note that $\mathcal{G}=\{1, 2, \cdot \cdot \cdot, n_i\}$ is the set of distributed generators, $\mathcal{E}=\{1, 2, \cdot \cdot \cdot, n_j\}$ is the set of power lines, and $\mathcal{N}=\{1, 2, \cdot \cdot \cdot, n_k\}$ is the set of power consuming loads.

\subsection{Dynamical model of DC Microgrids}
\begin{figure}[t]
    \centering
    \includegraphics[width=\columnwidth]{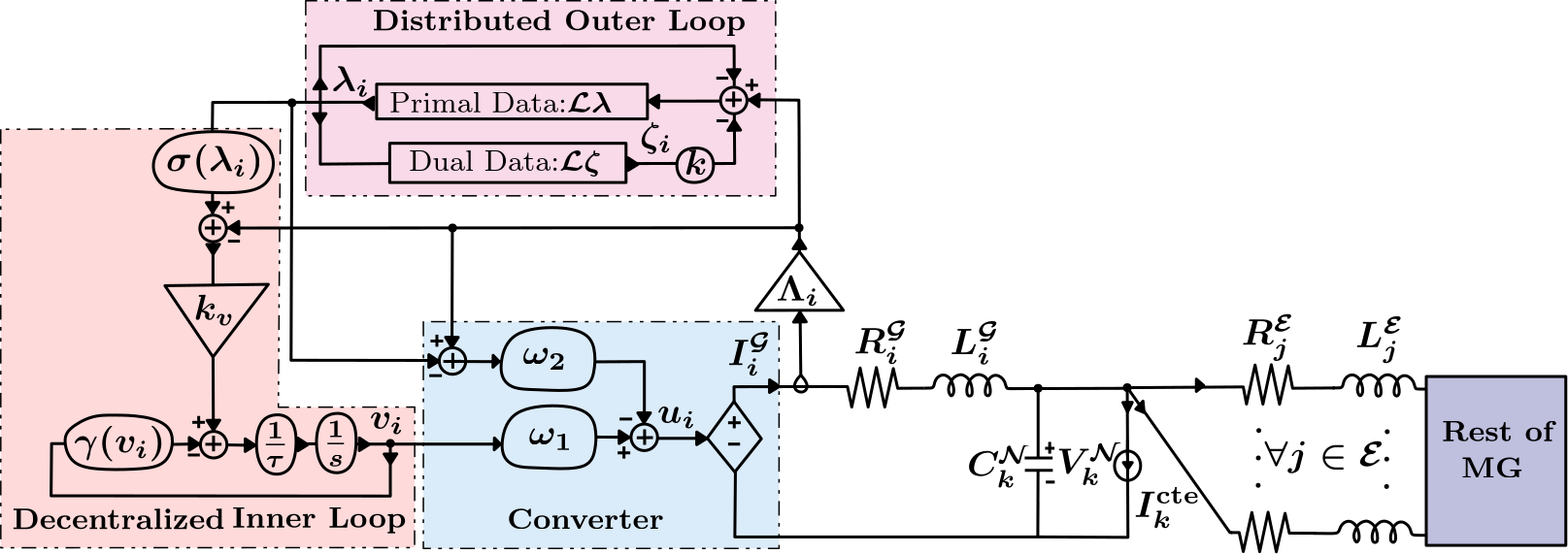}
    \caption{DC Microgrid dynamics - single unit perspective}
    \label{Image:TOT_EL}
\end{figure}

The architecture of DC microgrids is inherently multilayered, comprising a physical layer that interconnects agents and loads via RL-modeled power lines, and a cyber layer that facilitates information exchange among neighboring agents. The agents are the distributed generators (DGs), located close to the power-consuming loads (ZI-loads), and are effectively interfaced with the rest of the DC MG through voltage-controlled converters. The converters are considered equivalent zero-order models, hence, the internal voltage controller and associated dynamics are not considered in this paper. The DGs are interconnected both electrically and via distributed communication links, forming a cyber-physical grid. Graph theory is used to establish the physical and virtual interconnections, see Appendix A in \cite{Babak_AC} for the precise definitions of the included graphs. 
Following the model presented in Fig.~\ref{Image:TOT_EL}, the electrical dynamics of the $i$th DG, connected to the $j$th power line and the $k$th power-consuming load, are given in \eqref{eq:Physical-Layer}--derived from applying Kirchhoff's current and voltage law.
\begin{subequations}\label{eq:Physical-Layer}
\begin{IEEEeqnarray}{rCl}
L_i^\mathcal{G}\Dot{I}_i^\mathcal{G}&=&u_i - \sum_{k} b_{ik}^\mathcal{G}V_k^\mathcal{N}-R_i^\mathcal{G}I_i^\mathcal{G},\\
L_j^\mathcal{E}\Dot{I}_j^\mathcal{E}&=& -\sum_{k}b_{jk}^\mathcal{E}V_k^\mathcal{N}-R_j^\mathcal{E}I_j^\mathcal{E},\\
C_k^\mathcal{N}\Dot{V}_k^\mathcal{N}&=&\sum_{j}b_{kj}^\mathcal{E}I_j^\mathcal{E}+\sum_{i}b_{ki}^\mathcal{G}I_i^\mathcal{G}-G_k^\mathrm{cte}V_k^\mathcal{N}-I_k^\mathrm{cte},\label{eq:physical_load}
\end{IEEEeqnarray}
\end{subequations}
where $L_i^\mathcal{G}$, $R_i^\mathcal{G}$, $I_i^\mathcal{G}$, and $u_i$ are respectively the inductance, resistance, current, and voltage of the \textit{i}th DG. $L_j^\mathcal{E}$, $R_j^\mathcal{E}$, and $I_j^\mathcal{E}$ are respectively the inductance, resistance, and current of the \textit{j}th power line. $C_k^\mathcal{N}$, $V_k^\mathcal{N}$, $G_k^\mathrm{cte}$ and $I_k^\mathrm{cte}$ are respectively the shunt capacitance, its voltage, the constant conductance, and constant current of the \textit{k}th power-consuming load. The elements $b_{ik}^\mathcal{G}$ and $b_{jk}^\mathcal{E}$ correspond to the incidence matrices, characterizing arbitrary current flows within the network.

\subsection{Nonlinear Nested Distributed Control Framework}
Motivated by the previously proposed distributed controllers in \cite{Babak_Set_Point, Poppi_J1}, this paper proposes a modified version of the \emph{nonlinear} control framework, imposing two nested control loops: a decentralized(/primary) inner-loop controller that regulates current deviations with respect to a saturated set-point value, and a distributed(/secondary) outer-loop controller that restores operating conditions. The proposed control configurations aim to ensure simultaneous achievement of the following two control objectives:
\begin{itemize}[label=\textendash] 
    \item voltage containment within pre-specified limits,
    \item proportional current sharing in steady state.
\end{itemize}
The voltage containment objective is required to hold uniformly for all admissible initial conditions and disturbances.
These control objectives are mathematically formulated as follows: 
\begin{equation} 
    \begin{split}
        &V_\mathrm{min} \leq V_i^\mathcal{G}(t) \leq V_\mathrm{max}, \quad \forall \; t \in \; \mathbb{R}_{\geq 0},\\
        & \lim_{t \to \infty} \left( I_i^\mathcal{G}(t)/I_i^\mathrm{rated} -I_l^\mathcal{G}(t)/I_l^\mathrm{rated}\right)=0, \quad \forall i,l \in \mathcal{G}
    \end{split}
\end{equation}

\noindent where $V_i^\mathcal{G}$ and $I_i^\mathrm{rated}$ are respectively the voltage output and rated current of the $i$-th DG, and $V_\mathrm{min}$ and $V_\mathrm{max}$ are the minimum and maximum values of the voltage. The distributed controller in  \cite{Babak_Set_Point} was derived by means of the Lagrange multiplier method, where it is ensured that the system stabilizes at an equilibrium, which is the argument that minimizes a cost function. More precisely, the cost function was chosen such that current sharing is achieved in steady state when all units cooperatively decide on the optimal operating point, facilitated through communication among neighboring units. The distributed outer-loop controller and the decentralized inner-loop controller (local integrator for current regulation) aim to simultaneously ensure the two control objectives: voltage containment and proportional current sharing. 
However, the control proposal lacks stability guarantees. Thus, to facilitate obtaining a \emph{scalable} G.E.S. certificate for the proposed DC MG in this research, we adjust the structure of the control proposal in the following ways, resulting in \eqref{eq:Control-Layer}. First, we \emph{saturate} the proportional current set-point $\lambda$ appearing in the decentralized inner-loop controller \eqref{regulator} by means of a continuous strictly monotonically increasing (sigmoid) function $\sigma(\lambda)$, given in \eqref{sigmasat}. Second, we include the same saturation in the primal dynamics of the distributed outer loop, in \eqref{distributed_lambda}. Third, we update the leakage function of the regulator $ \rho(v)$ with the \emph{smooth} and continuous nonlinear function given in \eqref{leakage}. Finally, to support the industrial standard (adopting proportional-integral controllers (PI)), we have included a proportional controller ($\omega_2(\lambda_i, I_i^\mathcal{G})$) in our distributed framework, given in \eqref{omega2}. 
Note that the distributed optimizer relies on obtaining the rated current from each DG. Hence, the output of the physical network is then defined as the ratio of the DGs actual current to its rated current: $\Lambda_iI_i^\mathcal{G}=I_i^\mathcal{G}/I_i^\mathrm{rated}, \forall i \in \mathbb{R}^{n_i}$. 
For the \textit{i}th DG, the complete control scheme is then proposed as follows:
\begin{subequations} \label{eq:Control-Layer}
\begin{align}
    &u_i=V_i^{set}= \omega_1(v_i) - \omega_2(\lambda_i, I^\mathcal{G}_i),\label{omega}\\
    &\omega_1(v_i)=V^*+\Delta_1 \mathrm{tanh}\left(v_i/\Delta_1\right), \label{omega1}\\
    &\omega_2(\lambda_{i}, I^\mathcal{G}_i)=\Delta_2 \mathrm{tanh}\left(\mathrm{k}_p(\Lambda_{i}I_i^\mathcal{G}-\lambda_i)\right), \label{omega2}\\
    &\tau_{i} \Dot{v}_i=-\rho(v_i)v_i+\mathrm{k}_v(\sigma(\lambda_i)-\Lambda_{i}I_i^{\mathcal{G}}), \label{regulator}\\
    &\tau^p_{i} \Dot{\lambda}_i = \Lambda_iI_i^{\mathcal{G}}-\sigma(\lambda_i)- \sum_{j\in N_i} [a_{ij} (\zeta_i-\zeta_j)-\mathrm{k} a_{ij} (\lambda_j-\lambda_i)], \label{distributed_lambda}\\
    &\tau^d_{i} \Dot{\zeta}_i = \sum_{j\in N_i} a_{ij} (\lambda_i-\lambda_j), \label{distributed_zeta}\\
    &\sigma_i(\lambda_i)= \mathcal{K}_I+\Delta_I \tanh\left(\lambda_i/\Delta_I\right),\label{sigmasat}\\
    &\rho(v_i)=\alpha (1+0.5[\tan\mathrm{h}(\mathrm{b}(v_i-v_\mathrm{pos}))-\tan\mathrm{h}(\mathrm{b}(v_i-v_\mathrm{neg}))]).  \label{leakage}
\end{align}
\end{subequations}
The distributed control framework depends on three controller states where $\lambda_i$ and $\zeta_i$ are the communicated states of the cyber network with respective time constants $\tau_i^p$ and $\tau_i^d$, and $v_i$ is the decentralized controller state with associated time constant $\tau_i$. $u_i$ is the control actuator, constructed as a weighted sum of the two nonlinear functions $\omega_1(v_i)$ and $\omega_2(\lambda_i, I_i^\mathcal{G})$. For voltage containment, we define: $\mathrm{k}_v>0$ as the integrator gain, tuning the integration speed; $\mathrm{k}_p>0$ as the proportional gain; $V_\mathrm{max}=(1+\mu) V_\mathrm{n}$ and $V_\mathrm{min}=(1-\mu)V_\mathrm{n}$, for $\mu>0$, as the pre-specified voltage limitations from the networks nominal voltage, $V_\mathrm{n}$; $V^*=\frac{1}{2}(V_\mathrm{min}+V_\mathrm{max})$ as the central point within the safe voltage range; $\Delta=\frac{1}{2}(V_\mathrm{max}-V_\mathrm{min})$ as the maximum allowed deviation from $V^*$; and $\Delta_1$ and $\Delta_2$ are the weightings of the two functions given that $\Delta_1+\Delta_2= \Delta$. 
$\tanh(\cdot)$ is a hyperbolic tangent function used to saturate the argument of $\omega_1$ and $\omega_2$ if they exceed the predefined limits. $\tanh(\cdot)$ is also used in $\rho(v_i)$ as a nonlinear leakage expression operating as an anti-wind-up function when we reach voltage saturation under the given arguments; $v_\mathrm{pos}=\Delta\tanh^{-1}[(V_\mathrm{max}-v_\mathrm{tol}-V^*)/\Delta]$ and $V_\mathrm{neg}=\Delta\tanh^{-1}[(V_\mathrm{min}+v_\mathrm{tol}-V^*)/\Delta]$, where $v_\mathrm{tol}$ is the saturation tolerance, $\alpha$ is the leakage coefficient scaling the upper bound value, and $\mathrm{b}$ is implemented to scale the steepness of the curve. Moreover, we use $\mathrm{tanh}(\cdot)$ in $\sigma_i(\lambda_i)$ to bound the influence of the distributed outer-loop controller when entering the decentralized inner-loop with $\mathcal{K}_I=\frac{1}{2}(I^\mathrm{[p.u.]}_{\text{min}}+I^\mathrm{[p.u.]}_{\text{max}})$ and $\Delta_I=\frac{1}{2}(I^\mathrm{[p.u.]}_{\text{max}}-I^\mathrm{[p.u.]}_{\text{min}})$, where $I_\mathrm{max}=(1+\varphi)[\mathrm{p.u.}]$ and $I_\mathrm{min}=(1-\varphi)[\mathrm{p.u.}]$, for $\varphi \geq 0$. 
Finally, $a_{ij}$ is the \emph{adjacency matrix} element, describing the communicating DGs where $N_i$ is the set of the neighboring DGs in the cyber layer, and $k$ is a positive gain. 
\section{System Steady State and Stability Analysis}
\label{Stability}

In this section, we aim to prove that the equilibrium of \eqref{complete_sys_compact} is globally exponentially stable using singular perturbation theory and Lyapunov theory \cite{Khalil}. We start by expressing the system dynamics \eqref{eq:Physical-Layer}-\eqref{eq:Control-Layer} in the compact form:
%
\begin{subequations}\label{complete_sys_compact}
    \begin{align}
    L^\mathcal{G}\Dot{I}^\mathcal{G}&=\omega_1(v) - \omega_2(\lambda, I^\mathcal{G}) - \beta^\mathcal{G}V^\mathcal{N}-R^\mathcal{G}I^\mathcal{G} ,\label{complete_sys_compact_a}\\
    L^\mathcal{E}\Dot{I}^\mathcal{E}&= -\beta^\mathcal{E}V^\mathcal{N}-R^\mathcal{E}I^\mathcal{E}, \label{complete_sys_compact_b}\\
    C^\mathcal{N}\Dot{V}^\mathcal{N}&=\beta^{\mathcal{E}\top}I^\mathcal{E}+\beta^{\mathcal{G}\top}I^\mathcal{G}-G^\mathrm{cte}V^\mathcal{N}-I^\mathrm{cte},  \label{complete_sys_compact_c}\\
    \tau \Dot{v}&=-\gamma(v)+\mathcal{K}_v(\sigma(\lambda)-\Lambda I^\mathcal{G}), \label{complete_sys_compact_d}\\
    \tau_p \Dot{\lambda} &=\Lambda I^\mathcal{G}-\sigma(\lambda)-\mathcal{L}\zeta-k\mathcal{L}\lambda, \label{complete_sys_compact_e}\\
    \tau_d \Dot{\zeta} &= \mathcal{L}\lambda ,\label{complete_sys_compact_f}
    \end{align}
\end{subequations}

\noindent with $\mathcal{L}$ denoting the \emph{Laplacian} matrix,  containing the consensus properties of the communication network. All vectors and matrices in the above dynamics are defined in \emph{Appendix} \ref{App_1}.
\begin{assumption}
\label{ass_Laplacian}
    It is assumed that the communication network (represented by the Laplacian matrix $\mathcal{L}$) is strongly connected and undirected. 
\end{assumption}
\begin{assumption} (Time-Scale Separation)
\label{Asump_time_Const}
    It is assumed that the outer-loop time constants are greater than the largest time constant of the fast dynamics such that $\varepsilon \ll \tau_p \leq \tau_d$ or $\varepsilon \ll \tau_p,\tau_d$, given that $\varepsilon$ characterizes the slowest operation within the fast subsystem. 
\end{assumption}
\noindent Let \emph{Assumption} \ref{Asump_time_Const} hold. Thus, we assume sufficient time-scale separation between the electrical system and the inner-loop controller \eqref{complete_sys_compact_a}-\eqref{complete_sys_compact_d}, and the outer-loop controller \eqref{complete_sys_compact_e}-\eqref{complete_sys_compact_f}, given below in compact form as two systems with separated timescales.
\begin{align} \label{two_separated}
        \Dot{x}=s(x, z) \quad \text{and} \quad 
        \varepsilon \Dot{z}=f(x, z),
\end{align}
where $s(\cdot)$ and $f(\cdot)$ respectively denotes the slow and fast system dynamics, with $x=\mathrm{bcol}\{\lambda, \zeta\}\in \mathbb{R}^{n_\mathcal{S}}$ and $z=\mathrm{bcol}\{I^\mathcal{G}, I^\mathcal{E}, V^\mathcal{N}, v\} \in \mathcal{R}^{n_\mathcal{F}}$, where $n_\mathcal{S}=2 n_i$ and $n_\mathcal{F}=2\times n_i+n_j+n_k$.

\subsection{Equilibrium Analysis}\label{steady_state}

We derive the steady-state equations of \eqref{complete_sys_compact} by expressing $\dot x=0_{n_\mathcal{S}}$ and $\dot z=0_{n_\mathcal{F}}$. Hence, any steady state needs to satisfy
\begin{subequations}
\begin{IEEEeqnarray}{lCl}
        0_{n_\mathcal{S}}=s(\Bar{v}, \Bar{\lambda}, \Bar{\zeta},\Bar{I}^\mathcal{G}), \nonumber\\
        0_{n_\mathcal{F}}=f(\Bar{I}^\mathcal{G}, \Bar{I}^\mathcal{E}, \Bar{V}^\mathcal{N},\Bar{v}),\nonumber
\end{IEEEeqnarray}
\end{subequations}
where $(\bar \cdot)$ denotes the state value at its equilibrium. Let \emph{Assumption}~\ref{ass_Laplacian} hold, and by leveraging the property of the Laplacian matrix where $\mathcal{L}1_{n_i}=0_{n_i}$, the steady state operations of \eqref{complete_sys_compact_f} leads to all the DGs cooperatively agreeing upon one consensus value for the communicated state variable $\Bar{\lambda}=\lambda_s1_{n_i}$ where $\lambda_s\in\mathbb{R}$ is the singular consensus value and $1_{n_i}=\mathrm{col}(1)\in \mathbb{R}^{n_i}$. When considering steady state operation of \eqref{complete_sys_compact_d} without saturated voltages--inactivated leakage function $\rho(\Bar{v})$--  $\Lambda\Bar{I}^\mathcal{G}$ is forced equal to $\sigma(\bar \lambda)$. Subsequently, steady state operations provide consensus among the DGs in regards to the other communicated values when \eqref{complete_sys_compact_e} converge to the equilibrium and $\Bar{\zeta}=\zeta_s1_{n_i}$, where $\zeta_s$ is the singular consensus value. The steady state of the electrical network \eqref{complete_sys_compact_a}-\eqref{complete_sys_compact_c} is consequently uniquely defined. 

Furthermore, following the proof detailed in Section II.D in \cite{Babak_Set_Point}, our system attains \emph{optimal} current sharing operation when saturation is avoided. In this case, the controller simultaneously ensures proportional current sharing and voltage containment, provided that all neighboring DGs collaboratively determine the optimal set-point value $\bar \lambda$ such that $\Lambda\Bar{I}^\mathcal{G}=\sigma(\bar\lambda)$ at the equilibrium, and as long as $\rho(\bar \lambda)$ is uniformly defined for all DGs. However, in the case of voltage saturation, the DC MG is steered to another optimum that fails to meet the current sharing objective at the equilibrium due to the presence of the leakage function in \eqref{complete_sys_compact_d}. 

\subsection{Singular Perturbation Theory} \label{sec_SPT}
To accommodate the nested control loops, we first differentiate the dynamical system in \eqref{complete_sys_compact} into \emph{slow/fast} models in accordance with \emph{Assumption} \ref{Asump_time_Const}. Furthermore, under a stretched time-scale ($\mathbf{t}$), the singular perturbation problem is partitioned into the \emph{reduced}- and \emph{boundary layer} systems--based on the methodology in \cite{Khalil}. Subsequently, we conduct a Lyapunov stability analysis by proposing two separate Lyapunov candidates for the two time-scaled systems, aiming to prove G.E.S. of the individual systems. This facilitates finding a composite Lyapunov function for the singularly perturbed system as a weighted sum of the two Lyapunov candidates, used to conclude on G.E.S. of the \emph{full} system. 
    
\begin{assumption}\emph{(Frozen Variables in the Boundary Layer System)} \label{ass:frozen}
Let Assumption \ref{Asump_time_Const} hold, and let 
\begin{align*}
    \mathrm{h}(x)=\mathrm{bcol}\{\mathrm{h}^{I^\mathcal{G}}_i(x), \mathrm{h}_j^{I^\mathcal{E}}(x), \mathrm{h}^{V^\mathcal{N}}_k (x), \mathrm{h}^v_i(x)\}, \forall i \in \mathcal{G}, \forall j \in \mathcal{E}, \forall k \in \mathcal{N},
\end{align*}
be the unique solution of $\varepsilon \dot z=f(x, z)$ for $\varepsilon \approx 0$; thus, $\mathrm{h}^{I^\mathcal{G}}(x), \;\mathrm{h}^{I^\mathcal{E}}(x), \; \mathrm{h}^{V^\mathcal{N}}(x), \; \mathrm{h}^{v}(x)$ are defined by the solutions of the following equations respectively:
    \begin{IEEEeqnarray*}{lCl}\label{h(x)}
   \begin{bmatrix}
        \omega_1(\mathrm{h}^v(x))-\omega_2(\lambda, \mathrm{h}^{I^\mathcal{G}}(x))-\beta^\mathcal{G}\mathrm{h}^{V^\mathcal{N}}(x)-R^\mathcal{G}\mathrm{h}^{I^\mathcal{G}}(x)\\
        -\beta^\mathcal{E}\mathrm{h}^{V^\mathcal{N}}(x)-R^\mathcal{E}\mathrm{h}^{I^\mathcal{E}}(x)\\
        \beta^{\mathcal{G}\top}\mathrm{h}^{I^\mathcal{G}}(x)+\beta^{\mathcal{E}\top} \mathrm{h}^{I^\mathcal{E}}(x)-G^\mathrm{cte}\mathrm{h}^{V^\mathcal{N}}(x)-I^\mathrm{cte}\\
        -\rho(\mathrm{h}^v(x))+K_v(\sigma(\lambda)-\Lambda \mathrm{h}^{I^\mathcal{G}}(x))
    \end{bmatrix}=0_{n_\mathcal{F}}
    \end{IEEEeqnarray*}
Furthermore, when expressed in terms of incremental states, we define
    \begin{align*}
        \mathrm{H}(\tilde x)&\triangleq
        \mathrm{bcol}\{ \mathrm{h}(x)-\mathrm{h}(\bar x)\} 
        = \mathrm{bcol}\{\mathrm{h}(\tilde x+\bar x)-\mathrm{h}(\bar x)\} \\
        &= \mathrm{bcol}\{\hat{\mathrm{H}}(\tilde x)-\mathrm{h}(\bar x)\},
    \end{align*}
    with $x=\tilde x + \bar x$. Under Assumption~\ref{Asump_time_Const}, we assume that the slow states entering the fast dynamics can be treated as frozen variables, such that
    $x=\bar x$ and therefore $\tilde x = 0$, resulting in 
    $$\hat{\mathrm{H}}(\tilde x)=\mathrm{h}(\bar x) 
    \rightarrow \mathrm{H}(\tilde x)=\{\mathrm{h}(\bar x)-\mathrm{h}(\bar x)\}=0_{n_\mathcal{F}}.$$
    Moreover, let $\tilde y\triangleq \mathrm{bcol}\{\tilde z - \mathrm{H}(\tilde x)\}$ be the error between the actual fast dynamics $(\tilde z)$ and the quasi-steady state $(\mathrm{h}(\tilde x))$. Under sufficient time-scale separation, we assume $\bar z = \mathrm{h}(\bar x)$ such that $\bar y=0$ in the boundary layer system. Thus, from the above expressions we let $$\Tilde{\mathbf{z}}= (y-\bar y)+(\mathrm{h}(x)-\mathrm{h}(\bar x))=\tilde y,$$ when noticing that $y=\tilde y + \bar y$.
\end{assumption}

\begin{Prop5} \label{teorem_SPT} (Singular Perturbed Problem) Let Assumptions \ref{Asump_time_Const} and \ref{ass:frozen} hold. Consider the closed loop dynamics in \eqref{complete_sys_compact} and let 
\begin{align*}
\Omega_1(\tilde y^v) &\triangleq\omega_1(\tilde y^v+\mathrm{h}^v(\bar x))-\omega_1 (\mathrm{h}^v(\bar x)), \\
    \Omega_2( \tilde y^{I^\mathcal{G}}) &\triangleq \omega_2(\tilde y^{I^\mathcal{G}}+\mathrm{h}^{I^\mathcal{G}}(\bar x))-\omega_2(\mathrm{h}^{I^\mathcal{G}}( \bar x)), \\
    \Gamma(\tilde y^v) &\triangleq\gamma\left (\tilde y^v+\mathrm{h}^v(\bar x)\right)-\gamma\left (\mathrm{h}^v( \bar x)\right ), \\
    \Sigma(\tilde \lambda) &\triangleq \sigma(\lambda) - \sigma (\bar \lambda),\\
    \hspace{0.4in}\mathcal{Q}_s \triangleq\mathrm{bdiag}\{\tau_p, &\tau_d\}>0, \quad \mathcal{P}_s \triangleq \mathrm{bdiag}\{k\mathcal{L}, \mathcal{B}_\zeta \}>0,\\
    &\hspace{-0.9in}\mathcal{J}_s\triangleq\begin{bmatrix}
        \mathbf{0}_{n_i\times n_i} & -\mathcal{L}\\
        \mathcal{L} & \mathbf{0}_{n_i\times n_i}
    \end{bmatrix},\quad \kappa_1\triangleq \mathrm{bdiag}\{\mathbf{1}_{n_i\times n_i}, \mathbf{0}_{n_i\times n_i}\},\\
    &\hspace{-1.1in}\mathcal{Q}_f \triangleq\mathrm{bdiag}\{L^\mathcal{G},L^\mathcal{E},C^\mathcal{N} \}>0,\quad\mathcal{P}_f \triangleq \mathrm{bdiag}\{R^\mathcal{G},R^\mathcal{E},G^\mathrm{cte} \}>0,\\
    &\hspace{-0.4in}\mathcal{J}_f\triangleq\begin{bmatrix}
        \mathbf{0}_{n_i\times n_i} & \mathbf{0}_{n_i\times n_j} & -\beta^\mathcal{G}\\
        \mathbf{0}_{n_j\times n_i} & \mathbf{0}_{n_j\times n_j} & -\beta^\mathcal{E}\\
        \beta^{\mathcal{G}\top} & \beta^{\mathcal{E}\top} & \mathbf{0}_{n_k\times n_k}
    \end{bmatrix},\\
    &\hspace{-0.4in}\kappa_2\triangleq \mathrm{bdiag}\{\mathbf{1}_{n_i\times n_i},\mathbf{0}_{n_j\times n_j},\mathbf{0}_{n_k\times n_k}\},
\end{align*}
where $ \mathcal{B}_\zeta=\mathrm{diag}(\mathcal{B}^\zeta_i) \in \mathbb{R}^{n_i \times n_i}$ is an arbitrary small leakage term introduced in the $\zeta$ coordinate in \eqref{complete_sys_compact_f}. The two separated systems in \eqref{two_separated} can then be expressed as the singular perturbed problem in \eqref{SP_Compact_Inc} under the stretched timescale $\mathbf{t}=(t/\varepsilon)$, where $\hat{s}(\tilde x, \mathrm{H}(\tilde x))$ and $\hat{f}(\tilde y)$ represent the reduced and boundary layer systems, respectively. 
   \begin{subequations} \label{SP_Compact_Inc}
        \begin{IEEEeqnarray}{lCl}
            &\hspace{-0.55in}\hat{s}(\tilde x, \mathrm{H}(\tilde x)) : \big \{ 
             \mathcal{Q}_s \Dot{ \tilde x} = [\mathcal{J}_s-\mathcal{P}_s]\tilde x + \kappa_1[\Lambda\mathrm{H}^{I^\mathcal{G}}(\tilde x) - \Sigma(\tilde \lambda)], \label{RS_compact_inc}\\
            &\hspace{-.1in}\hat{f}( \tilde y): \label{BL_compact_inc}
            \begin{cases} 
                \mathcal{Q}_f \partial \tilde y_f/\partial \mathbf{t} = [\mathcal{J}_f-\mathcal{P}_f] (\tilde y_f)+ \kappa_2 [\Omega_1(\tilde y^v) - \Omega_2 ( \tilde y^{I^\mathcal{G}})],\\
                \tau \partial \tilde y^v/\partial \mathbf{t}= -\Gamma( \tilde y^v) - \mathcal{K}_v \Lambda \tilde y^{I^\mathcal{G}}- \mathcal{B}_v (\tilde y^v),
            \end{cases} 
        \end{IEEEeqnarray}
    \end{subequations} with $\tilde y_f\triangleq\mathrm{bcol}\{\tilde y^{I^\mathcal{G}}, \tilde y^{I^\mathcal{E}}, \tilde y^{V^\mathcal{N}}\}$.

\end{Prop5}
\begin{proof}
     First, we consider $\varepsilon$--characterized by the time constant associated with the slowest fast-dynamical state--to be significantly small such that the velocity of $\Dot{z}\propto (1/\varepsilon)$ behaves instantaneously fast. Thus, for $\varepsilon\approx 0$, the fast system in \eqref{complete_sys_compact_a}-\eqref{complete_sys_compact_d} quickly reaches a \emph{quasi-steady state} given the instantaneous fast dynamics $\mathrm{h}(x)=\mathrm{bcol}\{\mathrm{h}^b(x)\}, \forall b\in [I^\mathcal{G}, I^\mathcal{E}, V^\mathcal{N}, v]$. \footnote{In this system, the fast dynamics exclusively depend on the slow state $\lambda$. However, for the generality of this proof, we keep $\mathrm{h}(x)$ formulated as a function of all potential slow states.}  
    We further define $y=\mathrm{bcol}\{y^{I^\mathcal{G}}_i, y^{I^\mathcal{E}}_j, y^{V^\mathcal{N}}_k, y^v_i\}$ as the error between the \emph{actual} fast dynamics and the quasi-steady state: $y^b\triangleq \mathrm{bcol} \{z^b-\mathrm{h}^b(x)\}, \; \forall b\in [I^\mathcal{G}, I^\mathcal{E}, V^\mathcal{N}, v]$. Moreover, to facilitate later applying Lyapunov theory, we express the system states using their incremental variables $\tilde z= z - \bar z$, $\tilde y=y-\Bar{y}$ and $\tilde x=x-\Bar{x}$. Thus, when expressed in terms of incremental states, the \emph{actual} fast dynamics are given by $\Tilde{\mathbf{z}} \triangleq \tilde y + \mathrm{h}(\tilde x + \bar x) - \mathrm{h}(\bar x)=\tilde y + \mathrm{H}(\tilde x)$, and the dynamics of the two time-separated systems in \eqref{two_separated} are expressed as follows:
    \begin{align}\label{slow/fast_Compact_Inc}
       \begin{split}
        &s(\tilde x, \Tilde{\mathbf{z}}) : \begin{cases}
         \mathcal{Q}_s \Dot{ \tilde x} = [\mathcal{J}_s-\mathcal{P}_s]\tilde x + \kappa_1[\Lambda \Tilde{\mathbf{z}}^{I^\mathcal{G}} - \Sigma(\tilde \lambda)],
         \end{cases} \\
        &f(\tilde x, \Tilde{\mathbf{z}}) :  \begin{cases}
            \mathcal{Q}_f \Dot{\Tilde{\mathbf{z}}}_f = [\mathcal{J}_f-\mathcal{P}_f] \Tilde{\mathbf{z}}_f + \kappa_2 [\Omega_1(\tilde x, \Tilde{\mathbf{z}}^v) - \Omega_2 (\tilde x, \Tilde{\mathbf{z}}^{I^\mathcal{G}})],\\
            \tau \Dot{\tilde z}^v = -\Gamma(\tilde x, \Tilde{\mathbf{z}}^v) + \mathcal{K}_v [\Sigma(\tilde \lambda)-\Lambda \Tilde{\mathbf{z}}^{I^\mathcal{G}}] - \mathcal{B}_v \Tilde{\mathbf{z}}^v. 
        \end{cases} 
        \end{split}
    \end{align} 
    For compactness, we have defined the following for the slow model; the cyber inertia matrix $\mathcal{Q}_s$; the symmetrical cyber resistance matrix $\mathcal{P}_s$; and the interconnection matrix $\mathcal{J}_s$ and allocation matrix $\kappa_1$--with its dynamics previously defined in \emph{Theorem} \ref{teorem_SPT}. Moreover, for the fast model, we have defined the following; the reduced fast state vector $\Tilde{\mathbf{z}}_f\triangleq\mathrm{bcol}\{\Tilde{\mathbf{z}}^{I^\mathcal{G}}, \Tilde{\mathbf{z}}^{I^\mathcal{E}}, \Tilde{\mathbf{z}}^{V^\mathcal{N}}\}$, the physical inertia matrix $\mathcal{Q}_f$; the physical resistance matrix $\mathcal{P}_f$; and the interconnection matrix $\mathcal{J}_f$ and allocation matrix $\kappa_2$--dynamics defined in \emph{Theorem} \ref{teorem_SPT}.
    
    For mathematical purposes, we have introduced an additional leakage constant $\mathcal{B}_\zeta$, ensuring $\mathcal{P}_s$ to be a fully diagonal matrix. Note that $\mathcal{B}_\zeta$ does not need to compensate for any off-diagonal elements in the cyber resistance matrix. Hence, we choose the value to be arbitrarily small not to hinder the capability of achieving consensus for an optimal value of $\lambda$. Moreover, we have implemented an additional constant leakage $\mathcal{B}_v$ in the slow model to bring dissipation and improve the stability margins. These adjustments are further discussed in Section \ref{sec_Case_Studies}. 
    
    The time-separated systems in \eqref{slow/fast_Compact_Inc} are then expressed as the \emph{singular perturbation problem} in \eqref{SP_Compact_Inc}, partitioned into the reduced system \eqref{RS_compact_inc} and the boundary layer system \eqref{BL_compact_inc} divided under the stretched timescale $\mathbf{t}=(t/\varepsilon)$ where $t$ is the time when $\varepsilon \approx 0$. The slow model $s(\tilde x, \Tilde{\mathbf{z}})$ then instantaneously attains a quasi–steady state, since the fast dynamics are considered instantaneously fast, implying that $\tilde y\approx 0 $ in the reduced system \eqref{RS_compact_inc}. Furthermore, under \emph{Assumption}~\ref{ass:frozen}, the slow states entering the boundary layer are considered frozen variables, causing $\tilde x=0$ and $\bar y=0$, such that $\Tilde{\mathbf{z}}=\tilde y$ and $\Sigma(\tilde \lambda) = 0$ in the boundary layer system \eqref{BL_compact_inc}. 
    \end{proof}

\subsection{Lyapunov Stability Analysis} \label{sec_lyap}

\begin{Prop5} \emph{(Global Exponential Stability)} \label{GES_proof} 
Consider the singular perturbed problem in \eqref{SP_Compact_Inc}, and suppose that $\Omega_1(\tilde y^v)$, $\Gamma(\tilde y^v)$, and $\Omega_2(\tilde y^{I^\mathcal{G}})$ are Lipschitz and element-wise strictly monotonically increasing sigmoid functions in $\tilde y^v$ and $\tilde y^{I^\mathcal{G}}$ respectively. Moreover, let $$\mathcal{M}(\lambda) \triangleq \left[\sigma ( \lambda) - \Lambda \mathrm{h}^{I^\mathcal{G}}(\lambda)\right].$$ If $\mathcal{M}(\lambda)$ is strictly monotonically increasing; then, there exists $\varepsilon^*>0$ such that for all $\tau_p, \tau_d \gg \varepsilon^*$ the system in \eqref{complete_sys_compact} is globally exponentially stable under the following bounds:
\begin{align} 
    &-\tilde y^{{I^\mathcal{G}}\top} \Omega_2(\tilde y^{I^\mathcal{G}}) \leq - \nu ||\tilde y^{I^\mathcal{G}}||^2, \label{fast_Bound_1}\\
    &- \Omega_1(\tilde y^v)^\top [\mathcal{K}_v\Lambda]^{-1} [\Gamma(\tilde y^v)-\mathcal{B}_v\tilde y^v]\leq - \Phi ||\tilde y^v||^2, \label{fast_Bound_3}\\
    &-(\lambda -\Bar{\lambda})^\top [\mathcal{M}(\lambda)-\mathcal{M}(\Bar{\lambda})] < -\gamma ||\tilde \lambda||^2,\label{bound_3}
\end{align}
for some $(\nu,\Phi, \gamma) \; \in \mathbb{R}_{>0}$.
\end{Prop5}
\begin{proof} 
    Considering the singularly perturbed system in \eqref{SP_Compact_Inc}, we propose the subsequent Lyapunov candidates for the boundary layer system and the reduced system, to ensure global exponential stability of the two time-separated systems. First, we define the Lyapunov candidate for the boundary layer system as 
    \begin{align}\label{Lyap_fast}
    V_f(\tilde y)&=\frac{1}{2}\tilde y_f^\top \mathcal{Q}_f\tilde y_f + \int_{0}^{\tilde y^v}\Omega_1(\eta)^\top\tau[\mathcal{K}_v\Lambda]^{-1} \partial \eta >0.
    \end{align}
    Per \emph{Theorem} \ref{teorem_SPT}, $\mathcal{Q}_f>0$, thus ensuring the first quadratic term to be positive definite. When examining the integral dynamics--where $\mathcal{K}_v$ and $\Lambda$ are positive definite by definition--$\Omega_1(\cdot)$ is already defined in \emph{Theorem}~\ref{GES_proof} as a strictly monotonically increasing function in $\tilde y^v$. Consequently, when multiplied by the increment of the curve ($\eta$), the integral is ensured to be positive definite for any value of $\tilde y^v$, and the Lyapunov function is positive for $\tilde y \neq 0_{n_\mathcal{F}}$ and radially unbounded.
    Taking the time derivative of the Lyapunov function then gives
    \begin{align}
    \Dot{V}_f( \tilde y)&=\nabla^\top V_f(\tilde y)\Dot{\Tilde{y}} \label{derive_Lyap_Fast}\\
        &= - \tilde y_f^\top\mathcal{P}_f \tilde y_f \nonumber+ \tilde y^{{I^\mathcal{G}}\top} [\Omega_1(\tilde y^v) \nonumber - \Omega_2(\tilde y^{I^\mathcal{G}})] \nonumber\\
        &\hspace{0.14in}+ \Omega_1(\tilde y^v)^\top \tau[\mathcal{K}_v \Lambda]^{-1}\frac{\partial (\tilde y^v)}{\partial t} \nonumber\\
        &= - \tilde y_f^\top \mathcal{P}_f \tilde y_f \nonumber-\tilde y^{{I^\mathcal{G}}\top} \Omega_2(\tilde y^{I^\mathcal{G}} )\nonumber \\
        &\hspace{0.14in}- \Omega_1(\tilde y^v)^\top [\mathcal{K}_v\Lambda]^{-1} [\Gamma( \tilde y^v)+\mathcal{B}_v\tilde y^v]\leq 0, \nonumber
    \end{align}   
    where we have used the skew-symmetry of the interconnection matrix $\mathcal{J}_f$ in the first equality. The final inequality follows as the strictly monotonically increasing functions; $\Omega_1(\tilde y^v)$ and $\Omega_2(\tilde y^{I^\mathcal{G}})$ are positive definite when multiplied by their increments or when multiplied by another strictly monotonically increasing function of the same argument. Accordingly, we conclude upon global asymptotic stability (G.A.S.) of the boundary layer system under the stability bounds in \eqref{fast_Bound_1} and \eqref{fast_Bound_3}, determined solely by the norm of the fast dynamical states $\tilde y$.
    Furthermore, to guarantee G.E.S. we first need to find an upper bound of the Lyapunov function $V_f(\tilde y) \leq c||\tilde y||^2$. Then, by using the fact that $\Omega_1(\tilde y^v)$ adheres to the Lipschitz condition, we to contain the value of the integral as follows
\vspace{-.1in}
\begin{align}
    \int_{0}^{\tilde y^v}[\Omega_1(\eta)]^\top \partial \eta \leq \mathsf{L} \int_{0}^{\Tilde{y}^v} \eta  \;\; \partial \eta, \nonumber
\end{align}
for some scalar $\eta > 0$ and $\mathsf{L}>0$ is the Lipchitz constant. Thus, $V_f(\tilde y) \leq V_1(\tilde y)$ with $V_1(\tilde y) \triangleq \frac{1}{2}\tilde y_f^\top Q_f \tilde y_f + \frac{1}{2} \mathsf{L} \tilde y^{v}$ $= \frac{1}{2}\tilde y^\top \mathcal{Q}_f^*\tilde y\leq \Upsilon_\mathrm{max}(\mathcal{Q}_f^*)||\tilde y ||^2$ and  $\mathcal{Q}_f^*\triangleq \mathrm{bdiag}\{\mathcal{Q}_f, \mathsf{L}\}>0$ with $\Upsilon_\mathrm{max}$ being the maximum eigenvalue of $\mathcal{Q}_f^*$. Hence, we have 
\begin{align} \label{X1}
    V_f(\tilde y)\leq \Upsilon_\mathrm{max}(\mathcal{Q}_f^*)||\tilde y||^2.
\end{align}
Furthermore, the time derivative of the Lyapunov function along the trajectories of the fast system is bounded by 
\begin{align} \label{bound_layp_fast}
        \dot V_f(\tilde y)\leq -\Upsilon_\mathrm{min}(\mathcal{P}^*_f)||\tilde y||^2, 
\end{align}
for $\Upsilon_\mathrm{min}$ being the minimum eigenvalue of \eqref{PF^*} where we have incorporated \eqref{fast_Bound_1} and \eqref{fast_Bound_3} in \eqref{derive_Lyap_Fast} to define $\mathcal{P}_f^*$ as follows: 
\begin{align}\label{PF^*}
\mathcal{P}_f^*\triangleq\mathrm{bdiag}\{ (R^\mathcal{G}+\nu),\; R^\mathcal{E},\; G^\mathrm{cte}, \;  \Phi\}. 
\end{align}

Finally, using \eqref{X1} and \eqref{bound_layp_fast}, we conclude upon G.E.S. of the boundary layer system with a convergence rate given by $$\epsilon=2\frac{\Upsilon_\mathrm{min}(\mathcal{P}^*_f)}{\Upsilon_\mathrm{max}(\mathcal{Q}_f^*)}.$$
Considering the reduced system \eqref{RS_compact_inc}, the Lyapunov candidate and its time-derivative along the trajectories are defined as
\begin{subequations}\label{Lyap_and_deriv_slow}
    \begin{align}
        W_s(\tilde{x})&=\frac{1}{2}\Tilde{x}^\top\mathcal{Q}_s\Tilde{x}>0, \hspace{0.7in} \text{for} \quad\tilde x \neq 0_{n_\mathcal{S}}, \label{Lyap_slow}\\
        \Dot{W}_s(\Tilde{x})&=\nabla^\top W_s(\tilde{x})\Dot{\Tilde{x}}\nonumber\\
        &=-\tilde x^\top \mathcal{P}_s\tilde x -(\lambda-\Bar{\lambda})^\top[\mathcal{M}(\lambda)-\mathcal{M}(\Bar{\lambda})] \leq 0.  
        \label{derive_lyap_slow}
    \end{align}
\end{subequations}
Per \emph{Theorem} \ref{teorem_SPT}, $\mathcal{Q}_s$ is a positive definite matrix; thus, $W_s(\tilde x)$ is of quadratic form, positive for $\tilde x \neq 0_{n_\mathcal{S}}$ and radially unbounded. Furthermore, when assessing its time derivative, the last inequality in \eqref{derive_lyap_slow} is satisfied as $\mathcal{P}_s >0$, and when assuming that $\mathcal{M}(\lambda)$ is strictly monotonically increasing. Hence, we conclude that the reduced system is G.E.S. under the stability bound in \eqref{bound_3}. 

   
    Since both $V_f(\tilde y)$ and $W_s(\tilde{x})$ are positive definite and radially unbounded, we conclude G.E.S. of the reduced and the boundary layer system. Thus, the singularly perturbed system in \eqref{SP_Compact_Inc}, satisfies all conditions given in \emph{Theorem} 11.4 in \cite{Khalil} and \emph{Theorem} \ref{GES_proof} and the system in \eqref{complete_sys_compact} is globally exponentially stable as there exists $\varepsilon^*>0$ for all $\tau_p, \tau_d \gg \varepsilon^*$.

\end{proof}

\section{ Practical Guidelines for G.E.S. Guarantees}\label{sec:tuning}
In the general case, the stability condition in \eqref{bound_3} highly depends on $\mathcal{M}(\lambda)$ to be a strictly monotonically increasing function. Consequently, for the applicability of this stability result, we need to obtain sufficient parameter setting enforcing the following assumption to hold such that G.E.S. is guaranteed $\forall \; t \in \; \mathbb{R}_{\geq 0}$
\begin{assumption}\label{ass:mon}
    We assume $\mathcal{M}(\lambda)$ to be a strictly monotonically increasing function; that is, $\mathcal{M}(\lambda)$  satisfies the following condition 
    $$\frac{1}{2}\left[\nabla^\top \mathcal{M}(\lambda)+\nabla\mathcal{M}(\lambda)^\top\right]>0.$$
\end{assumption}

To verify \emph{Assumption} \ref{ass:mon}, we compute the symmetric part of $\frac{\partial\mathcal{M}(\lambda)}{\partial \lambda}$. If $\frac{\partial\mathcal{M}(\lambda)}{\partial \lambda}|_\mathrm{sym}>0$, then \emph{Assumption} \ref{ass:mon} holds. 
\begin{IEEEeqnarray}{lCl} \label{matrix}
        \left.\frac{\partial \mathcal{M}(\lambda)}{\partial \lambda}\right|_\text{sym} = \begin{bmatrix}
            \frac{\partial (\sigma(\lambda_1)-\Lambda \mathrm{h}^{I^\mathcal{G}}_1(\lambda))}{\partial \lambda_1} & \cdots & \frac{-\partial \Lambda \mathrm{h}^{I^\mathcal{G}}_1(\lambda)}{\partial \lambda_{n_i}} \\
            \vdots & \ddots & \vdots\\
            \frac{-\partial\Lambda \mathrm{h}^{I^\mathcal{G}}_{n_i}(\lambda)}{\partial \lambda_1} & \cdots & \frac{\partial (\sigma(\lambda)-\Lambda \mathrm{h}^{I^\mathcal{G}}_{n_i}(\lambda))}{\partial \lambda_{n_i}}
            \end{bmatrix} 
\end{IEEEeqnarray}  
Considering the expression in \eqref{matrix}, the above matrix is positive definite when the matrix-diagonal is dominant and each diagonal element is sufficiently positive. However, \emph{Remark} \ref{Remark:inter_dep} reveals that when the leakage function $\rho(v)$ is activated during voltage saturation, inter-dependencies emerge between neighboring units in a fully interconnected system. This complicates the search for closed expressions which in turn limits the straightforward analytical determination of specific bounds that guarantee positive definiteness of \eqref{matrix}.
\begin{Remark}\emph{(Electrical Inter-Dependencies in the Boundary Layer System})\label{Remark:inter_dep}
    Considering a fully interconnected system operating under voltage saturation, the activated leakage function, $\rho(v)$, introduces inter-dependencies between neighboring DGs and loads. When examining $0_{n_\mathcal{F}}=f(I^\mathcal{G}, I^\mathcal{E}, V^\mathcal{N}, v)$ we obtain following expression for $v$
    \begin{align}
        v=\omega_1^{-1}[(\beta^\mathcal{G}\theta\beta^{\mathcal{G}\top}+\mathcal{R}^\mathcal{G})I^\mathcal{G} + \beta^\mathcal{G}\theta I^\mathrm{cte}+ \omega_2(\lambda,I^\mathcal{G})], \label{inter_v}\\
        \text{with} \quad \theta=[\beta^{\mathcal{E}\top}\mathcal{R}^{\mathcal{E}^{-1}}\beta^\mathcal{E}+G^\mathrm{cte}]^{-1}. \nonumber
    \end{align}
This highlights that the value of $v$ is influenced by both incidence matrices $\beta^\mathcal{G}$ and $\beta^\mathcal{E}$,, which in turn causes $\rho(v_i)$ to depend on the cross-coupled current terms $I_i^\mathcal{G}$. In fact, $\beta^{\mathcal{E}\top}\mathcal{R}^{\mathcal{E}^{-1}}\beta^\mathcal{E}$ is indeed the \emph{non-diagonal} $Y_\mathrm{bus}$ of the electrical network, emphasizing the dependence on neighboring currents. Hence, when the leakage function is activated, each of the local instantaneous fast dynamics (with respect to the instantaneous currents; i.e.,  $\mathrm{h}^{I^\mathcal{G}}(\lambda)$) is a function of its neighboring $\lambda$-values; that is, \vspace{-.08in}
    $$\Lambda \mathrm{h}^{I^\mathcal{G}}_i(\lambda_1, \cdot \cdot \cdot, \lambda_n) \quad \forall i \in \mathbb R^{n_i}. \vspace{-.08in}$$ 
\end{Remark}
    
With this in mind, we introduce the following parameter setting guidelines; summarized in \emph{Remarks} \ref{Remark:tuning}-\ref{remark_Scheme5}, to
\begin{enumerate}
    \item weaken the electrical inter-dependencies, and to
    \item strengthen positive diagonal dominance in \eqref{matrix}.
\end{enumerate}

\begin{Remark} \emph{(Minimizing Electrical Inter-Dependencies)} \label{Remark:tuning}
    Tuning the leakage coefficient, $\alpha$, to a sufficiently small value minimizes the inter-dependencies of $\mathrm{h}^{I^\mathcal{G}}(\lambda)$ such that $\frac{\partial \mathcal{M}}{\partial \lambda}|_\text{sym}$ approximately becomes a diagonal matrix. Moreover, when assessing the dynamics of \eqref{inter_v}, it is notable that higher values of the load-conductances, $G^\mathrm{cte}$, reduces $\theta$, thereby causing less electrical inter-dependencies.
\end{Remark}

\begin{Remark}\emph{(Positive Diagonal Dominance)} \label{rem_tuning2}
    By construction, $\sigma(\lambda)$ is strictly monotonically increasing in $\lambda$, and the solution of $\varepsilon \dot z=0$ for $\mathrm{h}^{I^\mathcal{G}}(\lambda)$ gives a monotonically increasing function for some sufficiently small $\alpha$. However the subtraction $\partial (\sigma (\lambda)- \Lambda \mathrm{h}^{I^\mathcal{G}})/\partial \lambda$ is not guaranteed to be sufficiently positive. Thus, increasing the rated currents (i.e., reducing $\Lambda$) reinforces positive diagonal entries.
\end{Remark}

\begin{Remark}\emph{(Virtual Leakage in the Reduced System)} \label{remark_Scheme5}
    We introduce a virtual leakage in the dynamics of the primary communication state as
    $$\tau_p\dot \lambda=\Lambda I^\mathcal{G} - \sigma(\lambda)-\mathcal{L}\zeta-k\mathcal{L}\lambda - \mathcal{B}_\lambda(\lambda-\lambda^*),$$
    where $\lambda^*$ denotes the constant equilibrium value under normal operating conditions. This modifies the definition of $\mathcal{M}(\lambda)$ as 
    $$\mathcal{M}(\lambda) \triangleq \left[\sigma ( \lambda) - \Lambda \mathrm{h}^{I^\mathcal{G}}(\lambda) +\mathcal{B}_\lambda (\lambda - \lambda^*)\right],$$
    such that the diagonal entities of \eqref{matrix} are derived as $$\frac{\partial (\sigma(\lambda_i)-\Lambda \mathrm{h}^{I^\mathcal{G}}_i(\lambda))+\mathcal{B}_i^\lambda (\lambda_i-\lambda_i^*)}{\partial \lambda}. $$ 
    Thus, positive diagonal entries may be guaranteed for a sufficiently large $\mathcal{B}_\lambda>0$.
\end{Remark}

Alternatively, introducing a virtual leakage in the communication layer—as per Remark \ref{remark_Scheme5}—provides a sufficient means to \emph{practically ensure} positive diagonal dominance when the inter-dependencies are negligible and $\mathcal{B}_\lambda$ attains a significant positive value. It is also worth noting that introducing this leakage in the reduced system reinforces the diagonal entries without introducing any electrical inter-dependencies. However, our controller is designed to drive the system to a steady state where $\sigma(\bar \lambda)=\Lambda \bar I^\mathcal{G}$, to ensure proportional current sharing--as described in Section~\ref{steady_state}. Consequently, a significant virtual leakage will weaken the proportional current sharing objective forcing $\bar I^\mathcal{G}$ equal to $(\sigma(\bar \lambda) + \mathcal{B}_\lambda(\bar \lambda -\lambda^*))$ in steady state. Thus, under this strategy, achieving near-optimal proportional current sharing performance necessitates careful tuning of this leakage. 

\section{Case Studies} \label{sec_Case_Studies}
\begin{figure}[!t]
    \centering
    \includegraphics[width=.9\columnwidth]{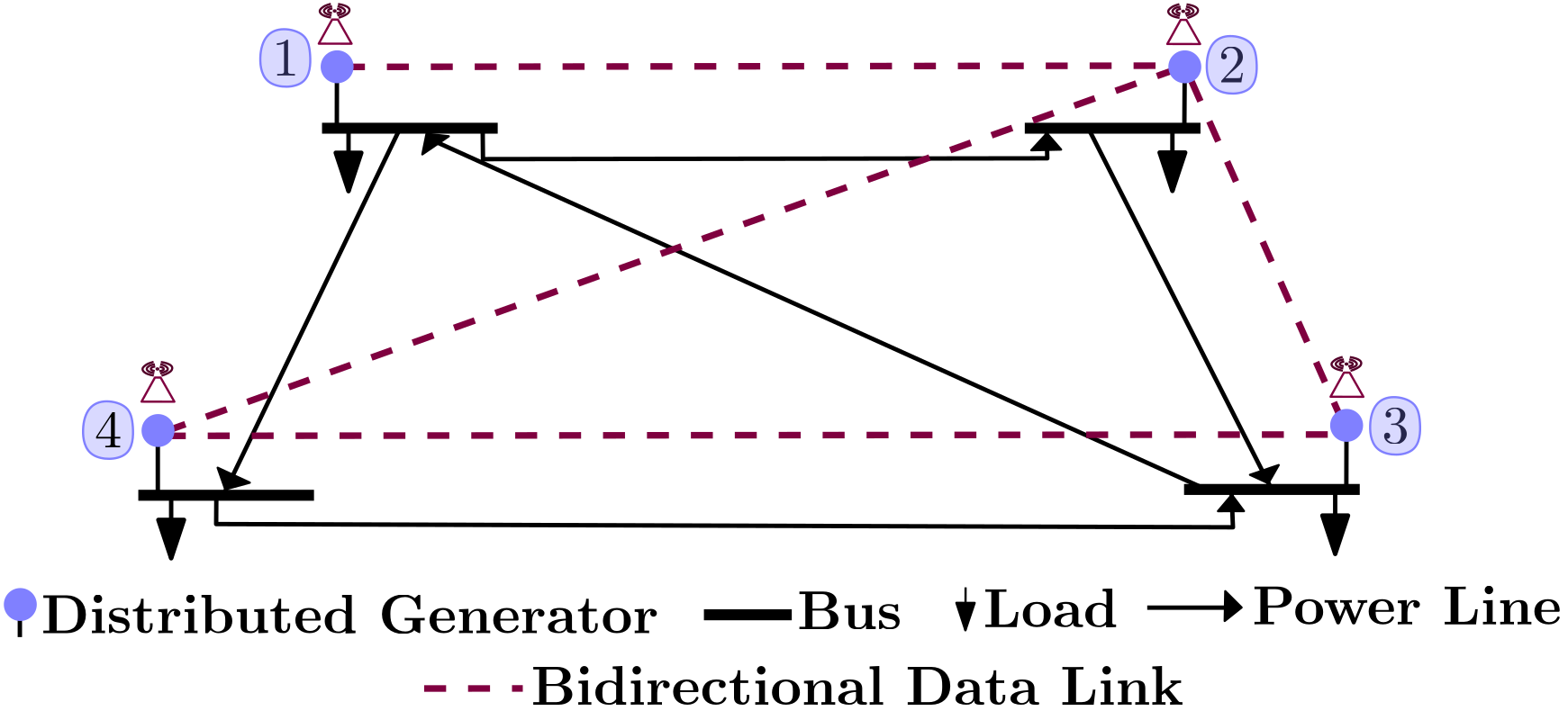}
    \captionsetup{font=small}
    \caption{Case Specific Microgrid}
    \label{Image:Sim_Overview}
\end{figure}

\begin{table}[!t]
    \centering 
    \captionsetup{font=small}
    \caption{Parameter values for case-specific MG} 
    \label{Table_Spec}
    \begin{tabular} {|c|c|c|c|c|}
    \hline
    \multicolumn{5}{|c|}{Generator specifications;  $ i \in \mathcal{G}$}\\
    \Xhline{0.95pt}
    $I_i^\mathrm{rated}$ $[\mathrm{A}]$ & 12 & 4 & 8 & 8\\
    \hline
    $R_i^\mathcal{G} [\mathrm{p.u}]$,$L_i^\mathcal{G} [\mathrm{p.u}]$  & 0.5 & 0.4 & 0.55 & 0.6\\
    \Xhline{0.95pt}
    \multicolumn{5}{|c|}{Load specifications; $k \in \mathcal{N}$}\\
    \Xhline{0.95pt}
    $C_k^\mathcal{N} [\mathrm{F}]$  & \multicolumn{4}{c|}{$2.2 \times 10^{-3}$}\\
    \hline
    $G_k^\mathrm{cte} [\Omega]$ & 1/40 & 1/30 & 1/30 & 1/30 \\
    \hline
    $I_k^\mathrm{cte} [\mathrm{A}]$ & 1 & 1.2 & 0.8 & 1 \\
    \Xhline{0.95pt}
    \multicolumn{5}{|c|}{Power lines specifications; $j \in \mathcal{E}$}\\
    \Xhline{0.95pt}
    \multicolumn{5}{|c|}{
        \begin{tabular}{c|c|c|c|c|c}
        $R_j^\mathcal{E} [\mathrm{p.u}]$, $L_j^\mathcal{E} [\mathrm{p.u}]$ & 1 & 2 & 2 & 1 & 1 \\
        \end{tabular}
        
    } \\
    \hline
    \end{tabular}
\end{table}

The control framework in \eqref{eq:Control-Layer} is tested by means of time-domain simulations in MATLAB/Simulink on both a 48-volt DC network admitting the dynamics in \eqref{eq:Physical-Layer} and on a modified version of the IEEE 33-bus benchmark test system. The 48-volt DC microgrid is powered by 4 DGs, interconnected electrically and through communication links according to the interconnection patterns depicted in Fig.~\ref{Image:Sim_Overview}. The specifications of the generators, loads, and power lines of the \emph{base-case} system; i.e., initially proposed electrical dynamics and control configurations prior to any stability-oriented modifications, are given in Table \ref{Table_Spec}, where all resistance ($R$) and inductance ($L$) values are specified in per unit (p.u) on a $(0.15 \Omega,\; 300 \mu H)$ base. The selected control parameters of the \emph{base-case} system are given below--selected to satisfy \emph{Assumption}~\ref{Asump_time_Const}-- with the time constants specified in seconds.
\vspace{-.06in}
\begin{equation} \label{init_cont}
\begin{aligned}
&\hspace{-.05in}\tau = \mathrm{diag}(1),  \hspace{-.07in}  & \tau_p &= \mathrm{diag}(10), \hspace{-.07in}  & \tau_d &=\mathrm{diag}(10), &\hspace{-.07in}\mathcal{K}_v &=\mathrm{col}(V^*),\\
&\hspace{-.05in}k = 10, & b  &= 5,  & \alpha&=\mathrm{diag}(V_\mathrm{max}), &\hspace{-.07in}K_p &=\mathrm{diag}(0).\\
\end{aligned}
\end{equation}
For voltage containment, the maximum allowed voltage deviation from the nominal voltage, $V_n=48V$, is $5\%$; $V_{\mathrm{max}}=1.05$[p.u] and $V_{\mathrm{min}}=0.95$[p.u]. Moreover, for the \emph{base-case} system we simulate the system without the proportional controller; i.e., relying solely on the decentralized integral controller, by selecting $\Delta_1=\Delta$ and $\Delta_2=0$.
\begin{figure*}[t!]
    \centering
    \includegraphics[width=\textwidth]{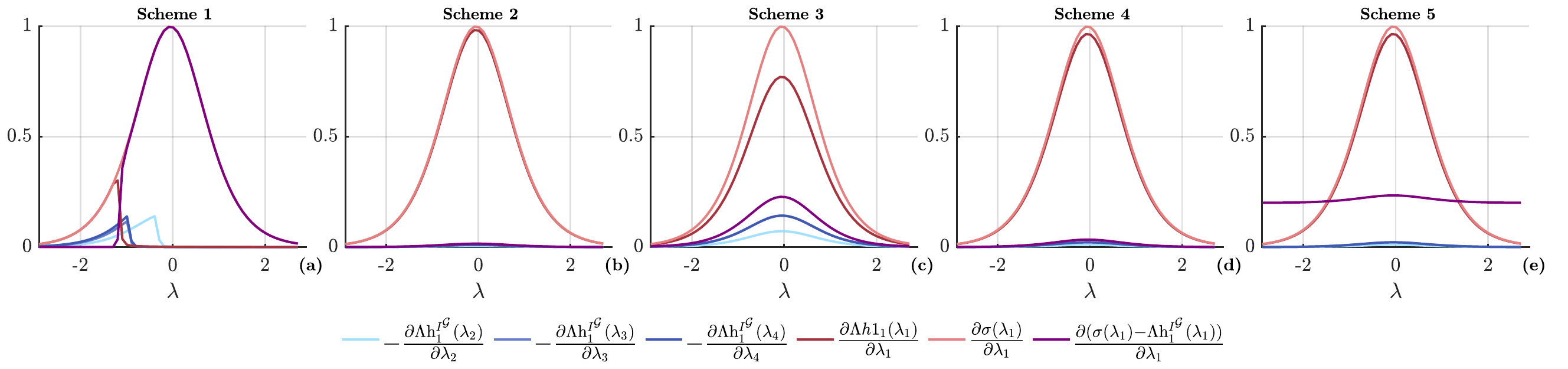}
    \captionsetup{font=small}
    \caption{Illustrative plots of selected components of $\frac{\partial M(\lambda)}{\partial \lambda}|_\mathrm{sym}$ over a range of $\lambda$-values}
    \label{Fig:Monoton_X3}
\end{figure*}
Throughout the subsequent case studies, we aim to find adequate tuning of our controller (and some equipment sizing) that satisfies \emph{Assumption}~\ref{ass:mon}, and further guarantee G.E.S. following the stability conditions in \emph{Theorem} \ref{GES_proof} under sufficient time-scale separation--satisfying \emph{Assumption}~\ref{Asump_time_Const}. Moreover, the applied case studies evaluate the control performance and ability to steer the DC MG toward (near-)optimal steady-state operation, while ensuring both voltage containment and proportional current sharing.

\subsection{Case Study 1: Practical Parameter Setting Schemes to Satisfy Assumption~\ref{ass:mon}} \label{CS1}
Preliminary, before testing the proposed control design on the case-specific DC microgrid, this case study examines some controller tunings (and electrical specifications) to identify the scenario that best satisfies \emph{Assumption}~\ref{ass:mon}, thereby ensuring G.E.S. conditions when testing the system. Section~\ref{sec:tuning} necessitates minimizing the electrical inter-dependencies introduced when the leakage function is activated, i.e., minimizing the off-diagonal entities of \eqref{matrix}: $\frac{\partial h^{I^\mathcal{G}}_i(\lambda_j)}{\lambda_j}$ while strengthening the diaconal entities: $\frac{\partial (\sigma(\lambda_i)- \Lambda_i \mathrm{h}_i^{I^\mathcal{G}}(\lambda_i))}{\partial \lambda_i}$. 
In Fig. \ref{Fig:Monoton_X3}, selected components of \eqref{matrix} are plotted over a range of $\lambda$-values under the following parameter setting schemes--consistent with the guidelines in Section~\ref{sec:tuning}--where $I^*$ and $G^*$ denote the \emph{base-case} values listed in Table~\ref{Table_Spec}.
\begin{IEEEeqnarray}{lCl}
    \begin{aligned} &\textbf{Scheme 1:}\hspace{.08in}\alpha=V_\mathrm{max},\hspace{.11in} I^\mathrm{rated}=I^*,\hspace{.32in}G^\mathrm{cte}=G^*,\hspace{.28in} \mathcal{B}_\lambda=0\\
    &\textbf{Scheme 2:}\hspace{.08in}\alpha=1e^{-11},\hspace{.08in} I^\mathrm{rated}=I^*,\hspace{.32in}G^\mathrm{cte}=G^*,\hspace{.27in} \mathcal{B}_\lambda=0\\
    &\textbf{Scheme 3:}\hspace{.08in}\alpha=1e^{-11},\hspace{.08in} I^\mathrm{rated}=5\times I^*,\hspace{.13in}G^\mathrm{cte}=5 \times G^*,\hspace{.09in} \mathcal{B}_\lambda=0\\
    &\textbf{Scheme 4:}\hspace{.08
    in}\alpha=1e^{-11},\hspace{.08in} I^\mathrm{rated}=1.5 \times I^*,\hspace{.04in}G^\mathrm{cte}=5 \times G^*, \hspace{.09in} \mathcal{B}_\lambda=0\\
    &\textbf{Scheme 5:}\hspace{.08
    in}\alpha=1e^{-11},\hspace{.08in} I^\mathrm{rated}=1.5 \times I^*,\hspace{.04in}G^\mathrm{cte}=5 \times G^*,\hspace{.09in} \mathcal{B}_\lambda=0.2
    \end{aligned}\nonumber
    \end{IEEEeqnarray}

Fig.~\ref{Fig:Monoton_X3}(a) shows that when $\alpha = V_\mathrm{max}$, \emph{Assumption}~\ref{ass:mon} fails to hold, as the diagonal entries are not uniformly positive definite and fails to dominate the off-diagonal entries. Consequently, G.E.S. cannot be guaranteed under Scheme~1. In contrast, Fig.~\ref{Fig:Monoton_X3}(b) indicates that under Scheme~2 the off-diagonal entries become negligible; however, a similar attenuation is also observed in the diagonal. Therefore, to satisfy \emph{Assumption}~\ref{ass:mon}, we further evaluate the controller on a system with higher load conductances and rated currents following \emph{Remarks}~\ref{Remark:tuning} and \ref{rem_tuning2}. Accordingly, Fig.~\ref{Fig:Monoton_X3}(c) depicts the selected entries under Scheme~3, where the diagonal entries remain continuously positive and are larger than the off-diagonal entries. However, increasing the rated currents strengthens the diagonal entries while simultaneously amplifying the off-diagonal entries, and thus, prohibiting a definitive conclusion regarding diagonal dominance. To address this, we propose Scheme~4--response shown in Fig.~\ref{Fig:Monoton_X3}(d)--which identifies a specific set of system parameters that yields negligible inter-dependencies and \emph{increases the likelihood} of having positive definite diagonal dominance---notice that this is not formally verified due to lack of closed expressions for $\mathcal{M}(\lambda)$.

By contrast, Scheme~5, shown in Fig.~\ref{Fig:Monoton_X3}(e), modifies Scheme~4 by incorporating a virtual leakage in accordance with \emph{Remark}~\ref{remark_Scheme5}. The resulting response \emph{practically ensures} a strictly positive diagonal significantly larger than the negligible off-diagonal entries, though at the cost of reducing the current sharing objective. Consequently, both Schemes~4 and~5 may satisfy \emph{Assumption}~\ref{ass:mon}, thereby providing sufficient conditions to guarantee G.E.S. for the considered DC microgrid.

\begin{Remark}
To broaden the applicability this papers result, a more general controller tuning can be attained by first minimizing the inter-dependencies--as per Scheme~2--and subsequently enhancing diagonal dominance via $\mathcal{B}_\lambda$, thereby satisfying Assumption~\ref{ass:mon} without relying on any electrical specifications. However, further simulations studies indicate that solely decreasing the leakage function $\rho(v)$ enforces momentary DG voltage saturation due to persistently elevated $v$-values.
Thus, to properly illustrate the system response under the proposed control schemes, we propose Scheme~5 as a modification of Scheme~4.
\end{Remark}

\subsection{Case Study 2: Control Performance under Stability Constraints} \label{CS2}

\begin{figure*}[t!]
    \centering
    \includegraphics[width=\textwidth]{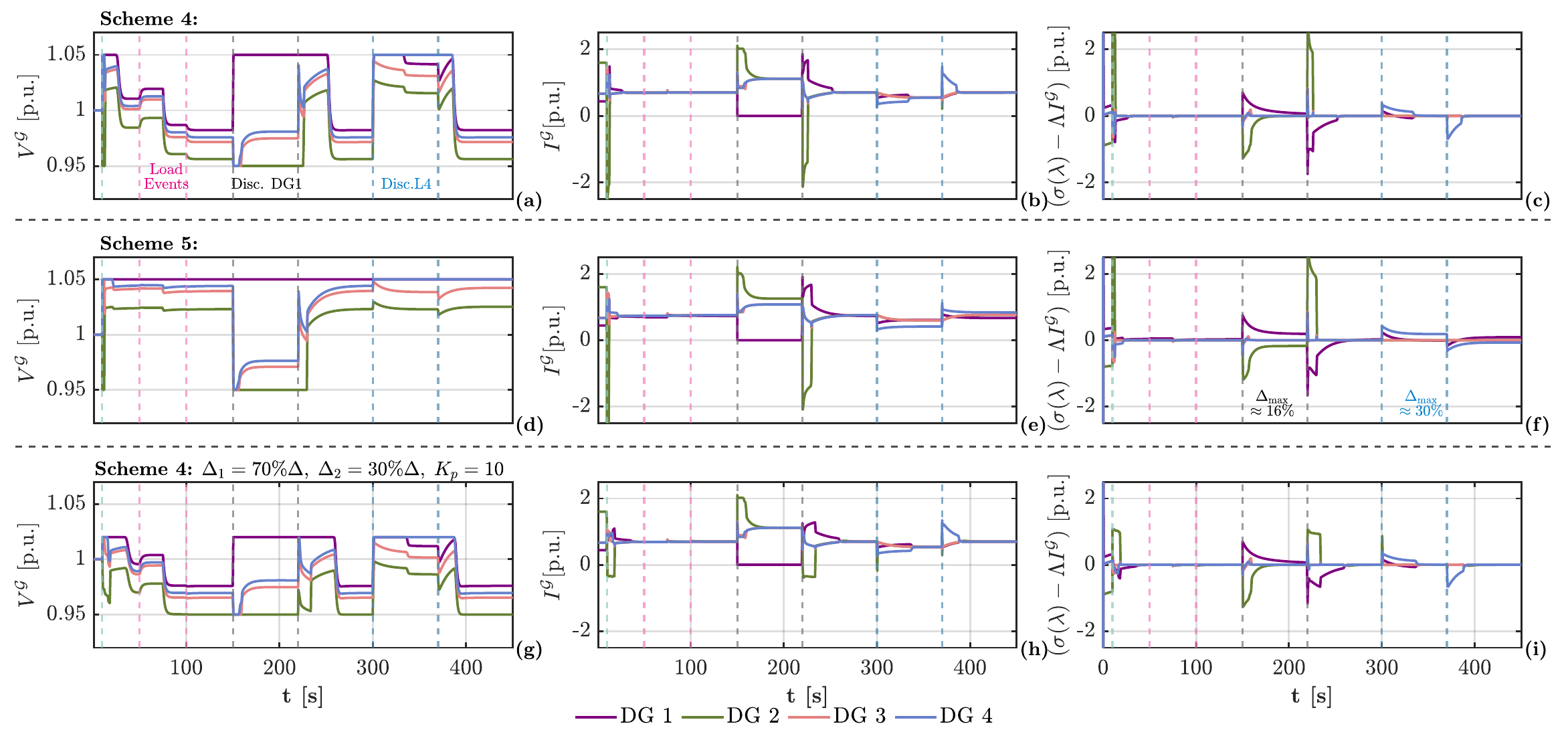}
    \captionsetup{font=small}
    \caption{System response under varying control parameter settings; Scheme 4 (a)-(c) (first row), Scheme 5 (d)-(f) (second row), Scheme 4 with PI-controller (g)-(i) (third row); (a), (d), (g) DG voltages; (b), (e), (h) generated currents; (c), (f), (i) integration errors}
    \label{Fig:Simumations_X9}
\end{figure*}

In this case study, we evaluate the performance and effectiveness of the proposed control framework in \eqref{complete_sys_compact} when applied to the case-specific DC microgrid under the parameter settings of Schemes~4 and~5, with the remaining system specifications provided in Table~\ref{Table_Spec} and \eqref{init_cont}. Additionally, we select $\mathcal{B}_\zeta= 1e^{-3}$ and $\mathcal{B}_v=1e^{-5}$ to improve the stability margins for smaller time-scale separations--further discussed in Section \ref{CS3}. Our objective is to verify the controllers ability to consistently ensure voltage containment and convergence to a steady state in which (near-)optimal proportional current sharing is achieved, under a range of operating conditions including both smaller load events and a set of plug-and-play scenarios, imposed to test the controllers robustness and validate its scalable properties. All simulated events are given in Table~\ref{Table_events}.  
\begin{table}[h!]
    \centering 
    \captionsetup{font=small} 
    \caption{Imposed System Events} \label{Table_events}
    {\huge 
    \resizebox{0.95\columnwidth}{!}{
        \begin{tabular}{|c|c|c|c|c|c|c|}
        \hline
          & $t=10\,\text{s}$ & $t=50\,\text{s}$ & $t=75\,\text{s}$ & $t=100\,\text{s}$ 
          & $t \in [150,220]\,\text{s}$ & $t \in [300,370]\,\text{s}$ \\ 
        \Xhline{1pt}
         Events 
         & \makecell{Control \\ activation} 
         & Inc. $I_1^{\mathrm{cte}}$ by $25\%$ 
         & Inc. $G_3^{\mathrm{cte}}$ by $15\%$ 
         & Red. $I_1^{\mathrm{cte}}$ by $15\%$ 
         & Disc. DG1 
         & Disc. Load 4 \\
        \hline
        \end{tabular}}
    }
\end{table}

Figs.~\ref{Fig:Simumations_X9}(a) and (d), illustrate that the controller ensures voltage containment and safely converge to stable operations across all imposed events. When employing Scheme~4, the voltages reach their maximum or minimum saturation levels only when DG1 and Load~4 are disconnected; however, they remain saturated within acceptable limits. In comparison, under Scheme~5, the DGs reach voltage saturation more rapidly due to the virtual leakage increasing the consensus value of $\lambda$, which in turn elevates the integral controller state which accelerates voltage saturation. Nevertheless, this does not pose any operational issues, as the voltages remain within acceptable limits.

When considering the proportional current sharing objective, the ideal system should exhibit a response where the DGs converge to a steady state in which their integration errors approach zero. This behavior is indeed observed in Fig.~\ref{Fig:Simumations_X9}(c) under Scheme~4. However, as discussed in Section~\ref{sec:tuning}, implementing the virtual leakage--as in Scheme~5--weakens the proportional current-sharing objective. From a practical perspective, we allow for a $\pm 5$\% deviation from \emph{optimal} proportional current sharing. However, the response in Fig.~\ref{Fig:Simumations_X9}(f) shows that under this Scheme, the integration error does not converge to zero in steady state, and actually deviates by 16\% during the disconnection of DG1 and by 30\% during the disconnection of Load~4. Consequently, \emph{practical} proportional current sharing is not achieved under significant topology changes, thus, we employ Scheme~4 in the following case studies.


Finally, in Figs.~\ref{Fig:Simumations_X9}(g)–(i), we present the system response under the widely used proportional–integral controller while maintaining the parameter settings of Scheme~4. The theoretical stability analysis in Section~\ref{Stability} accounts for this proportional component, and the depicted results are obtained by setting $K_p = 10$ and updating the actuator weightings to reflect the combined influence of the voltage containment function $\omega_1$ and the proportional controller $\omega_2$. Specifically, we assign $\Delta_1$ to contribute 70\% of $\Delta$ and $\Delta_2$ the remaining 30\%, where $\Delta$ denotes the allowable voltage deviation range.

As observed in Fig.~\ref{Fig:Simumations_X9}(g), the voltages remain within the acceptable range and achieve stable operations under all simulated events; however, as the arguments of $\omega_1$ and $\omega_2$ differ, they do not reach their saturation limits simultaneously, and the voltages do not reach $V_\mathrm{max}$ under these operating conditions. In some scenarios however, both functions may reach saturation simultaneously, which motivates defining $\Delta = \Delta_1 + \Delta_2$, thereby ensuring that the voltages are never forced beyond their allowable limits under any operations. When considering the currents in Figs.~\ref{Fig:Simumations_X9}(b) and (e), and integration errors in Figs.~\ref{Fig:Simumations_X9}(c) and (f), transient and sharp current peaks emerge when DG1 and Load 4 are disconnected. However, with the proportional controller activated the system response in Figs.~\ref{Fig:Simumations_X9}(h) and (l) illustrates a significant reduction in these current peaks, thereby providing more robust and reliable system performance. Furthermore, Fig.~\ref{Fig:Simumations_X9}(l) demonstrates that this modified PI controller maintains the proportional current-sharing objective, illustrating the ideal behavior in which the integration error converges to zero in steady state under all imposed events.

\begin{figure*}[t!]
    \centering
    \includegraphics[width=\textwidth]{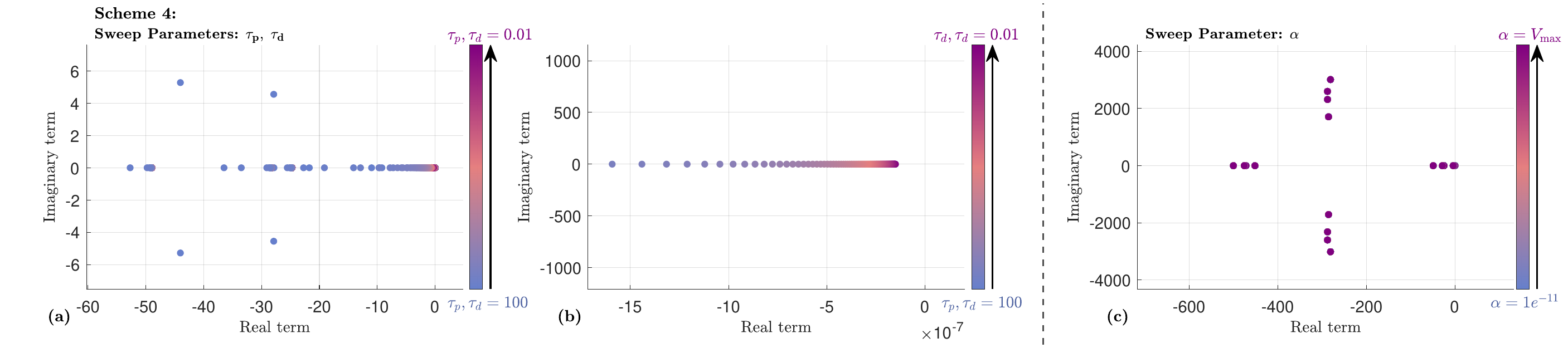}
    \captionsetup{font=small}
    \caption{Eigenvalues; (a) and (b) eigenvalues for a parametric sweep of $\tau_p, \tau_d$, (b) detailed image around zero; (c) eigenvalues for a parametric sweep of $\alpha$}
    \label{Fig:SmallSignal_X3}
\end{figure*}
\begin{figure*}[!t]
    \centering
    \includegraphics[width=\textwidth]{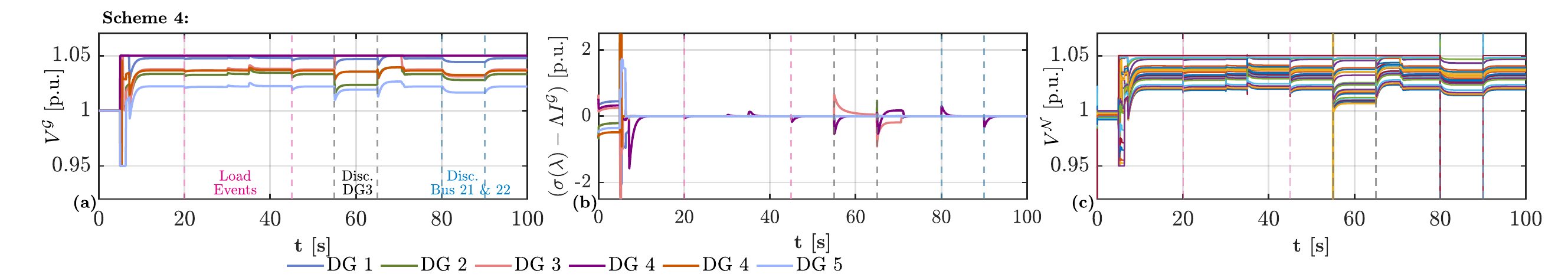}
    \captionsetup{font=small}
    \caption{IEEE 33-bus test system response: (a) generator voltages; (b) integration errors; (c) bus voltages}
    \label{Fig:Sim_IEEE}
\end{figure*}
\begin{figure}[!t]
    \centering
    \includegraphics[width=.9\columnwidth]{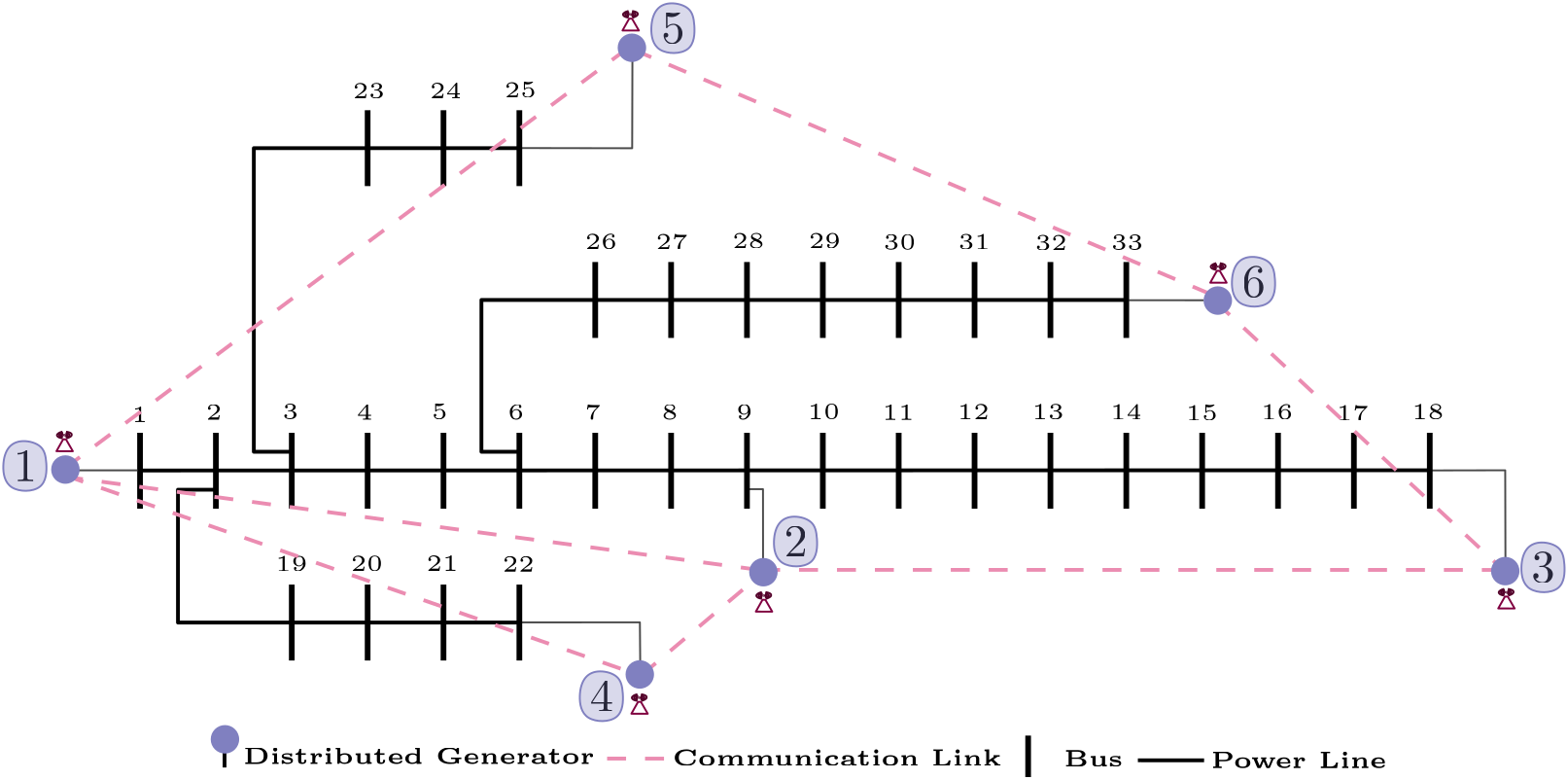}
    \captionsetup{font=small}
    \caption{ Modified IEEE 33-Bus radial distribution system}
    \label{Fig_IEEE}
\end{figure}

\subsection{Case Study 3: Small-Signal Stability Analysis} \label{CS3}
To complement the large-signal stability results, we conduct a small-signal stability analysis to obtain less conservative stability bounds and broaden the applicability of our findings. Given that large-signal stability is preserved with substantial margins--characterized by significant time-scale separation between the nested control loops and careful selection of parameter settings--we employ this small-signal eigenvalue analysis to further examine how the system operates under reduces large-signal stability margins to accommodate faster communication rates.

We begin by noting that the outer-loop dynamics depend on the cyber-network topology defined by the Laplacian matrix, which inherently yields a zero eigenvalue due to its mathematical properties. Hence, to properly evaluate the closed-loop systems small-signal stability, a small constant leakage ($\mathcal{B}_v$) is included in \eqref{complete_sys_compact_d} to shift the zero eigenvalue further into the left-half plane. Moreover, in Section~\ref{sec_SPT}, the negligible virtual leakage $\mathcal{B}_\zeta$ included in \eqref{complete_sys_compact_f} to add dissipation in the $\zeta$ coordinate, yet also serves a similar purpose. Thus, through both $\mathcal{B}_v$ and $\mathcal{B}_\zeta$, we are able to shift the zero eigenvalues associated to the Laplacians to the left-half plane, allowing us to assess small-signal stability of our system. That being said, it is important that these leakages remain sufficiently small to \emph{practically} guarantee consensus, such that the system converges to a \emph{near}-optimal equilibrium where $\lambda_i \approx \lambda_j$ and $\zeta_i \approx \zeta_j$. Accordingly, we select $\mathcal{B}_v = 1\times 10^{-5}$ and $\mathcal{B}_\zeta = 1\times 10^{-3}$; the resulting equilibrium values for $\lambda$ and $\zeta$ are detailed in Table~\ref{Table_equilibrium}, validating that this tuning ensures optimal consensus from a practical viewpoint.

\begin{table} [h!]
    \centering 
    \captionsetup{font=small}
    \caption{Equilibrium when $\mathcal{B}_\zeta=1\times 10^{-3}$ and $\mathcal{B}_v=1\times 10^{-5}$} \label{Table_equilibrium}
    \resizebox{0.95\columnwidth}{!}{
    \begin{tabular} {|c|c|c|c|c|c|c|c|} 
    \hline
     $\lambda_1$ & $\lambda_2$ & $\lambda_3$ & $\lambda_4$ & $\zeta_1$ &$\zeta_2$ &$\zeta_3$ &$\zeta_4$\\ 
    \Xhline{1pt}
     -0.3100 & -0.3100 & -0.3100 & -0.3100 & 0.00 & 0.00 & 0.00 & 0.00\\
    \hline
    \end{tabular}
    }
\end{table}
To examine how faster communication rates influence the system stability, we compute the eigenvalues of the case-specific DC microgrid for a parametric sweep of the communication time constants $\tau_p,\tau_d$. The resulting eigenvalue trajectories are given in \ref{Fig:SmallSignal_X3}~(a)(b) which illustrates that the convergence rates of the closed-loop system states approach zero when increasing the communication rates; however the system remains stable for all $\tau_p,\; \tau_d$. From a practical viewpoint, the system remains stable under small-signal disturbances, though with reduced stability margins. Thus, faster communication may allow for increased area of application, but requires practitioners to rely more on additional studies to compensate for the lack of large-signal stability guarantees, though, e.g., additional time-domain simulations. 

Furthermore, as the large-signal stability certificate strongly depends on tuning the leakage function to negligible values, we examine the eigenvalues for varying values of the leakage coefficient $\alpha$ to assess whether the small-signal stability results exhibit a similar dependence. The resulting eigenvalues are illustrated in Fig. \ref{Fig:SmallSignal_X3}~(c), revealing that the leakage function--and therefore the electrical inter-dependencies--has negligible impact on small-signal stability as the eigenvalues are kept constant during the parametric sweep of $\alpha$. This results from linearizing the system around the equilibrium point of the case-specific DC microgrid--under parameter setting Scheme 4 and with the selected values of $\mathcal{B}v$ and $\mathcal{B}\zeta$--during \emph{normal} operation; that is, following the topology shown in Fig.~\ref{Image:Sim_Overview} without load variations or unit disconnections. At this converged equilibrium, the voltages remain within saturation limits, and consequently, the leakage function remains inactive. However, Fig.~\ref{Fig:Simumations_X9}(a) illustrates that the DG voltages of the case-specific DC MG reach voltage saturation--thus, activating the leakage function--upon disconnecting DG1 and Load~4. Consequently, Fig. \ref{Fig:SmallSignal_X3}~(c) indicates that small-signal stability analysis alone does not provide sufficient means to guarantee stability in these scenarios. This reinforces the strength of the large-signal stability analysis as it provides robust \emph{practical} guidelines to consistently guarantee stability even when units are connected or disconnected.

\subsection{Case Study 4: Enhanced IEEE 33-Bus Benchmark Test System}\label{CS5}

In this case study, we evaluate the applicability of our controller by assessing its performance on the 12.66kV IEEE 33-bus radial distribution system in \cite{IEEE_33Bus}. The system- and communication network topology is given in Fig.~\ref{Fig_IEEE}, which follows the design in \cite{IEEE_33Bus}; however, the electrical system has been modified in the following ways.
\begin{itemize}
    \item Since the system in \cite{IEEE_33Bus} is originally AC, all specifications of the load buses and power lines are converted to DC parameters using the following standard conversion:
    \begin{align*}
    &G_i^\mathrm{load}[\mathrm{S}]=\frac{S_{i,\mathrm{load}}[\mathrm{VA}]}{V_n^2[\mathrm{V}^2]}=\frac{\sqrt{P_{i,\mathrm{load}}^2+Q_{i,\mathrm{load}}^2}}{V_n^2},\\
    &L_{i, \mathrm{PL}}[\mathrm{H}]=\frac{X[\Omega]}{2 \pi f[\mathrm{Hz}]}, \qquad \text{with} \;\; f=60[\mathrm{Hz}].
    \end{align*}
    \item We connect six DC DGs on buses 1, 9, 18, 22, 25, and 33, with their inherent dynamics given in \eqref{complete_sys_compact_a}. 
    \item Motivated by a similar case study in \cite{Babak_AC}, the rated capacities of the respective AC DGs were selected to be 3, 1, 0.75, 0.75, 0.75, and 1 MVA to meet the maximum load requirement when testing the system without any shunt capacitance. Accordingly, the rated currents 
    of the considered DC DGs are obtained using the following conversion: 
    \begin{align*}
        I_i^\mathrm{rated}[\mathrm{A}]=\frac{S_i^\mathrm{rated}[\mathrm{VA}]}{V_n[\mathrm{V}]}.
    \end{align*}
\end{itemize}
The system is tested under similar imposed events as in the previous simulations, where we impose load events for $t\in[20, 45]$; specifically, we increase the load consumption at buses 1, 12, 13, 28 by 25\% for $t\in[20, 30]$ and decrease the load consumption at buses 8 and 23 by 50\% for $t\in[35, 45]$. Further, we disconnect DG3 for $t\in[55, 65]$s, and disconnect buses 21 and 22 for $t\in[80,90]$s. Moreover, to satisfy the time-scale separation we first approximate the time-constants of the electrical system to be defined by the ratio of inertia to damping, and subsequently tune the rates of the controller to satisfy \emph{Assumption} \ref{Asump_time_Const}; thus, we set $\tau=\mathrm{diag}(0.1),\; \tau_p=\mathrm{diag}(1), \; \tau=\mathrm{diag}(1)$ while keeping the control settings of Scheme~4 \footnote{In the considered IEEE 33-bus system, the converters are modeled as voltage-controlled (zero order) models. Thus, inclusion of a more accurate dynamical model necessitates updating the time-scales accordingly to meet \emph{Assumption} \ref{Asump_time_Const}. }.

Fig.~\ref{Fig:Sim_IEEE}(a) demonstrates stable and satisfactory operations of the IEEE 33-bus system under the proposed control strategy. The DG voltages are maintained within safe operating ranges throughout all imposed events. Furthermore, Fig.~\ref{Fig:Sim_IEEE}(b) confirms that the communicating DGs achieve consensus, effectively driving the system to a steady state in which proportional current sharing is ensured, as evidenced by the integral error converging to zero during and after all events. In addition, Fig.~\ref{Fig:Sim_IEEE}(c) indicates that all bus voltages operate at acceptable values and exhibit stable behavior. These results verify that, even for this larger-scale test system, the proposed controller guarantees large-signal stability and supports scalable operations, allowing reliable plug-and-play connection and disconnection of units.

\section{Conclusion and Further Work}
\label{conc}
In conclusion, this paper establishes a \emph{scalable} global exponential stability (G.E.S) certificate for a DC microgrid with distributed control configurations. Advancing the result originally introduced in \cite{Poppi_SEST}, the proposed \emph{modified} distributed control framework relies on nonlinear nested control loops, designed to drive the system to a steady state that satisfies proportional current sharing, while enforcing voltage containment at all times.
To accommodate the nested control loop, we employ \emph{singular perturbation theory} and Lyapunov theory, to establish a G.E.S. certificate. Accordingly, stability is only guaranteed when some conditions are met, including sufficient time-scale separation at the border between the decentralized inner-loop dynamics and the distributed outer-loop controller. 
Specifically, the outer-loop controller (and inherent communication technologies) are assumed to operate at a slower time-scale than the rest of the system. 

The complexity of transcendental functions and multidimensional expressions makes it challenging to derive analytical bounds in which stability is guaranteed. To address this, we instead provide some practical parameter setting guidelines to ensure the stability conditions. While these adjustments enhance stability, they also modify the system equilibrium. Consequently, as the controller is originally designed to drive the system toward a steady state achieving proportional current sharing, these parameter setting schemes necessitate a careful characterization of the equilibrium to guarantee exponential convergence to the \emph{desired} operating point. Moreover, the selection of the parameter setting scheme should be guided by the intended objectives: a configuration that \emph{practically ensures} G.E.S., albeit at the expense of reduced proportional current-sharing performance, or a configuration which demands certain electrical specifications and relies on \emph{practical approximations}, though, guaranteeing desired proportional current sharing in steady state.

The stability results and control performance are validated on a case-specific DC microgrid under appropriate parameter-setting schemes. The results demonstrate stable operation with generator voltages maintained within predefined limits and convergence to \emph{near-optimal} proportional current sharing in steady state under all imposed system events--including generator and load disconnections--thereby highlighting the scalable nature of the proposed framework. Moreover, the simulations illustrate that adding a proportional channel improves the control performance by reducing unwanted current peaks. The stability is also analyzed from a small-signal perspective, demonstrating that for smaller system events, stability is guaranteed under less conservative time-scale separation requirements. Indeed, operation outside the large-signal stability conditions may still be acceptable under normal operations, specially if compensated by additional studies, such as extended time-domain simulations. Finally, the applicability of proposed control framework is tested on a modified large-scale benchmark system to validate its scalability and versatility, with simulation results demonstrating stable and optimal operations in accordance with the control objectives.

Finally, we note that our results rely on the assumption of zero-order converter models common in the power and control systems literature, and linear loads. Indeed, this assumption may in some cases prove insufficient, and more accurate converter models that take into account, e.g., nonlinearities or additional filtering and switching mechanisms should be used instead,
as well as extending the results to consider nonlinear constant power loads. We leave this natural extension to our research as a promising topic for future work.



\balance
\bibliographystyle{IEEEtran.bst}
\bibliography{IEEEabrv,Refs}

\appendices

\section{Mathematical Definitions}
\begin{equation*}
\begin{aligned}
L^\mathcal{G}&=\mathrm{diag}(L_i^\mathcal{G})\in \mathbb{R}^{n_i\times n_i}   & \hspace{0.4in}
I^\mathcal{G}&=\mathrm{col}(I_i^\mathcal{G}) \in \mathbb{R}^{n_i}\\ 
R^\mathcal{G}&=\mathrm{diag}(R_i^\mathcal{G}) \in \mathbb{R}^{n_i\times n_i}
 &   \omega_1(v)&=\mathrm{col}(\omega_1(v_i)) \in \mathbb{R}^{n_i}\\  \beta^\mathcal{G}&=[b_{ik}^\mathcal{G}]\in \mathbb{R}^{n_i\times n_k} & \omega_2(\lambda,I^\mathcal{G})&=\mathrm{col}(\omega_2(\lambda_i,I^\mathcal{G}_i) \in \mathbb{R}^{n_i}\\
 L^\mathcal{E}&=\mathrm{diag}(L_j^\mathcal{E})\in \mathbb{R}^{n_j\times n_j} &
I^\mathcal{E}&=\mathrm{col}(I_j^\mathcal{E}) \in \mathbb{R}^{n_j} \\
R^\mathcal{E}&=\mathrm{diag}(R_j^\mathcal{E}) \in \mathbb{R}^{n_j\times n_j} &
\beta^\mathcal{E}&=[ b_{jk}^\mathcal{E} ] \in \mathbb{R}^{n_j\times n_k}  \\
C^\mathcal{N}&=\mathrm{diag}(C_k^\mathcal{N})\in \mathbb{R}^{n_k\times n_k} &
V^\mathcal{N}&=\mathrm{col}(V_k^\mathcal{N}) \in \mathbb{R}^{n_k} \\
G^\mathrm{cte}&=\mathrm{diag}(G_k^\mathrm{cte}) \in \mathbb{R}^{n_k\times n_k} &
I^\mathrm{cte}&=\mathrm{col}(I_k^\mathrm{cte}) \in \mathbb{R}^{n_k} \\
v &=\mathrm{col}(v^c_{i}) \in \mathbb{R}^{n_i} &
\tau&=\mathrm{diag}(\tau_i) \in \mathbb{R}^{n_i\times n_i} \\
\gamma(v)&=\mathrm{col}(\rho(v_i)v_i)\in \mathbb{R}^{n_i \times n_i} & \sigma(\lambda)&=\mathrm{col}(\sigma(\lambda_i))\in \mathbb{R}^{n_i} \\ \mathcal{K}_v&=\mathrm{diag}(\mathrm{k}_v)\in \mathbb{R}^{n_i\times n_i} & \lambda&=\mathrm{col}(\lambda_i) \in \mathbb{R}^{n_i} \\
\Lambda&=\mathrm{diag}(1/I^\mathrm{rated}_{i}) \in \mathbb{R}^{n_i\times n_i} & \zeta&=\mathrm{col}(\zeta_i) \in \mathbb{R}^{n_i}\\
\tau_p&=\mathrm{diag}(\tau^p_i) \in \mathbb{R}^{n_i \times n_i} & \tau_d&=\mathrm{diag}(\tau^d_i) \in \mathbb{R}^{n_i \times n_i} \\ 
\mathcal{L}&=[l_{ij}]\in \mathbb{R}^{n_i \times n_i}, & \mathcal{B}_\zeta&=\mathrm{diag}(\mathcal{B}^\zeta_i) \in \mathbb{R}^{n_i \times n_i}\\
\mathcal{B}_v&=\mathrm{diag}(\mathcal{B}^v_i) \in \mathbb{R}^{n_i \times n_i}, & \mathcal{B}_\lambda&=\mathrm{diag}(\mathcal{B}^\lambda_i) \in \mathbb{R}^{n_i \times n_i}
\end{aligned}
\end{equation*}

\label{App_1}

\end{document}